\documentclass[8.5pt,twoside,twocolumn]{article}
\usepackage[super,sort&compress,comma]{natbib} 
\usepackage[version=3]{mhchem}
\usepackage{balance}
\usepackage{times,mathptm}
\usepackage{sectsty}
\usepackage{graphicx} 
\usepackage{lastpage}
\usepackage[format=plain,justification=raggedright,singlelinecheck=false,font=small,labelfont=bf,labelsep=space]{caption} 
\usepackage{fancyhdr}
\usepackage[english]{babel}
\usepackage{setspace}
\usepackage{amssymb}
\usepackage{amsmath}
\usepackage{booktabs}
\usepackage{color}
\usepackage{longtable}
\usepackage{midfloat}
\usepackage{url}

\begin{document}

\thispagestyle{plain}
\fancypagestyle{plain}{
\renewcommand{\headrulewidth}{1pt}}
\renewcommand{\thefootnote}{\fnsymbol{footnote}}
\renewcommand\footnoterule{\vspace*{1pt}%
\hrule width 3.4in height 0.4pt \vspace*{5pt}} 
\setcounter{secnumdepth}{5}

\makeatletter 
\renewcommand\@biblabel[1]{#1}            
\renewcommand\@makefntext[1]%
{\noindent\makebox[0pt][r]{\@thefnmark\,}#1}
\makeatother 
\renewcommand{\figurename}{\small{Fig.}~}
\sectionfont{\large}
\subsectionfont{\normalsize} 

\fancyfoot{}
\fancyhead{}
\renewcommand{\headrulewidth}{1pt} 
\renewcommand{\footrulewidth}{1pt}
\setlength{\arrayrulewidth}{1pt}
\setlength{\columnsep}{6.5mm}
\setlength\bibsep{1pt}

\twocolumn[
  \begin{@twocolumnfalse}
\noindent\LARGE{\textbf{Self-assembly scenarios of patchy colloidal particles}}
\vspace{0.6cm}

\noindent\large{\textbf{G\"unther Doppelbauer,$^{\ast}$\textit{$^{a}$} Eva G.~Noya,\textit{$^{b}$}
Emanuela Bianchi,\textit{$^{a}$} and Gerhard Kahl\textit{$^{a}$}}}\vspace{0.5cm}


 \end{@twocolumnfalse} \vspace{0.6cm}

  ]

\noindent\textbf{  The rapid progress in precisely designing the surface decoration of patchy
  colloidal particles offers a new, yet unexperienced freedom to
  create building entities for larger, more complex structures in soft
  matter systems. However, it is extremely difficult to \textit{predict} the
  large variety of ordered equilibrium structures
  that these particles are able to
  undergo under the variation of external parameters, such as
  temperature or pressure. Here we show that, by a novel combination
  of two theoretical tools, it is indeed possible to predict the
  self-assembly scenario of patchy colloidal particles: on one hand, a
  reliable and efficient optimization tool based on ideas of evolutionary
  algorithms helps to identify the ordered equilibrium structures to
  be expected at $T = 0$; on the other hand, suitable simulation
  techniques allow to estimate via free energy calculations  
  the phase diagram at finite temperature. With these powerful approaches we
  are able to identify the broad variety of emerging self-assembly scenarios
  for spherical colloids decorated by four patches and we investigate and discuss the
  stability of the crystal structures on modifying in a controlled way the tetrahedral arrangement of the patches.}
\section*{}
\vspace{-1cm}

\footnotetext{\textit{$^{a}$~Institut f\"ur
  Theoretische Physik and Center for Computational Materials Science
  (CMS), Technische Universit\"at Wien, Wiedner Hauptstra{\ss}e 8-10,
  A-1040 Vienna, Austria. E-mail: guenther.doppelbauer@tuwien.ac.at}}
\footnotetext{\textit{$^{b}$~Instituto de Qu{\' \i}mica F{\' \i}sica
  Rocasolano, CSIC, Calle Serrano 119, E-28006 Madrid, Spain.
  E-mail: eva.noya@iqfr.csic.es}}

Experimental \textit{and} theoretical investigations have provided
unambiguous evidence that colloids with chemically or
physically patterned surfaces (commonly known as ``patchy particles'') are
very promising mesoscopic entities that can be used in hierarchical
self-assembly processes to build up colloidal super-structures~\cite{Patchy_revExp,Patchy_rev2011}.
The anisotropy in the interactions of such particles in combination with the limited
functionality and selectivity of the bonds offer unlimited
possibilities for self-assembly scenarios. Thus, patchy
particles are celebrated ``to become the elementary brick of
tomorrow's self-assembled materials''~\cite{Sciortino-Kagome}, with
promising applications in photonic crystals, drug-delivery,
electronics~\cite{Kretz2009}, biomaterials, or
catalysis~\cite{Granick-Kagome}.

The possibility to tailor the interactions
of patchy particles almost deliberately represents the basis for
``bottom-up'' strategies which allow to build up materials with desired 
properties, starting from adequately designed units.
Suitable experimental
techniques~\cite{Mohwald2006,Kraft2009,Velev2010,VelevKretz} allow to
position patches on the colloidal surface and to define their spatial
extent with high precision.
A very impressive example for such an approach is a recent work on triblock Janus 
particles~\cite{Granick-Kagome}: after decorating colloids with two
hydrophobic caps of tunable area, particles self-organize in the two-dimensional Kagome lattice target structure. Complementary computer simulations~\cite{Sciortino-Kagome,RomanoSSoftMatter11}
have provided a complete phase diagram of the system,
including also the disordered, fluid phase.

For the case of triblock Janus particles the self-assembly scenarios
were easy to ``guess''. However, for more complex 
decorations and three dimensional systems, it is considerably more
difficult to identify \textit{all} ordered structures into which the
particles might self-assemble. Semi-empirical approaches,
applied over many years in hard matter physics, rely on a pre-selection
of candidate structures, based on experience, intuition 
or plausible arguments. In view of the rich wealth of unexpected ordered structures
in soft matter systems such a procedure is bound to fail.
In this contribution we propose a novel approach which helps to \textit{predict}
with high reliability the ordered equilibrium structures of a system of patchy
colloidal particles. We evaluate the phase diagram 
for different tetrahedral patch arrangements based on a highly reliable pre-selection step, 
and we investigate the stability of the emerging structures. We provide a comprehensive classification of these structures in dependence of the patch geometry and interpret them in terms of the strong competition 
between energetic, entropic and packing contributions.

We use a standard model for patchy particles,
introduced by Doye \textit{et al.}~\cite{DoyeLLANWKL07},
which has been used in numerous studies (e.g.,
Refs.~\cite{NoyaVDL07,NoyaVDL10,WilberDLNMW07,WilberDLL09,WilliamsonWDL11,vanderLindenDL11,JPCM10,JPCM11}).
An isotropic Lennard-Jones (LJ) potential (specifying the
spherical colloids) is modulated by an orientationally dependent
factor; the latter mimics the patches which can be ``located'' on
arbitrary positions and with variable extent on the surface of the
spherical colloidal particle. We use the LJ parameters
$\sigma$ and $\epsilon$ as units for length and energy.
As in previous work \cite{NoyaVDL10}, the patch width is fixed by the parameter $\sigma_{\rm pw}=0.3$, leading to rather narrow patches -- the interaction strength of a bond decreases to half its value when the patch alignment deviates by an angle of 20 degrees from the perfect one.  Additionally, the potential is truncated and smoothed at a cutoff distance of $r_{\rm cut}=2.5\sigma$ in order to prevent long-range interactions that are very unrealistic for colloids.
Usually colloidal interactions are even of much shorter range, but this choice allows us to perform a comparison with previous results, if available. We also note, that the long-range contributions (which we define by $r>1.5 r_{\rm min} \sim 1.68 \sigma$, where $r_{\rm min}$ is the minimum of the LJ potential) to the binding energies of the structures discussed in the following are well below ten percent of the total binding energy and of the same order of magnitude in all cases. In this contribution we consider only one value of the interaction range; however it has been shown, that a change in the interaction range can affect the regions of stability of the emerging phases \cite{NoyaVDL10,Romano10}.
Packing fractions $\eta$ are calculated as
$\eta=(\pi/6) (N \sigma^3 / V)$, $V$ and $N$ being the volume of
and the number of particles in the primitive cell, respectively.
Binding energy and pressure values are calculated as dimensionless quantities ($U^\star=U/(N\epsilon)$ and $P^\star=P\sigma^3/\epsilon$, respectively).

\begin{figure}[htbp]
\begin{center}
\includegraphics[width=8.0cm]{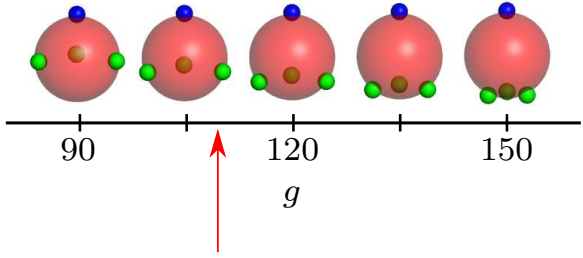}
\end{center}
  \caption{(Colour online) Visualization of the decoration of the patchy particles
    (large spheres) by the four patches (small spheres) as
    the geometric parameter $g$ varies. Here (and in the following
    figures) blue spheres specify the patch located on the
    north pole of each particle ($\mathbf{p}_1$), while green spheres specify the
    remaining three patches ($\mathbf{p}_k$). The arrow represents
    the $g$ value for which the patches form a regular tetrahedron.}
  \label{fig:geometry}
\end{figure}

Here we focus on
four-patch particles: introducing a geometrical parameter $g$, we 
vary the patch positions on the surface, ranging from a rather
elongated to a flat, compressed tetrahedral arrangement. More specifically, the first
patch $\mathbf{p}_1$ is fixed at the north pole of the particle, while the three
 remaining patches $\mathbf{p}_k$ (with $k \in \{2,3,4\}$) are located
at a degree of latitude related to $g$:
\begin{eqnarray*}
\mathbf{p}_1&=&\frac{\sigma}{2}\begin{pmatrix}0\\0\\1\end{pmatrix}\\
\mathbf{p}_k&=&\frac{\sigma}{2}\begin{pmatrix}\cos(2k \pi /3)\sin(g \pi/180)\\\sin(2k \pi /3) \sin(g \pi/180)\\\cos(g \pi/180)\end{pmatrix}
\end{eqnarray*}
See Figure \ref{fig:geometry} for a visualization.

Our investigations display for the first time the surprisingly broad variety of ordered equilibrium structures that the particles are able to form when the
regular tetrahedral arrangement of the patches is modified via the parameter $g$.

In order to investigate the self-assembly scenarios of the system, we use a two-step
approach, combining two efficient and reliable numerical tools.
First, we identify at \textit{vanishing} temperature $T$ possible 
ordered equilibrium structures by applying an optimization tool which is based on
ideas of evolutionary algorithms (EAs) and searches essentially among all
possible lattice structures~\cite{GottwaldKL05}. Second, we use suitably
designed computer simulations~\cite{VegaN07, NoyaVDL10, review} to evaluate
the free energy of the
candidate structures previously identified, as well as the free energy of disordered phases, leading
to the phase diagram at \textit{finite} $T$.
By performing full free energy calculations,  we are thus including entropic contributions 
at nonvanishing temperature.

In numerous applications to a wide variety of systems, EAs
have turned out to be reliable, efficient and robust~\cite{WoodleyC08,
OganovG06, BandowH06, JPCM10}: they cope well with high dimensional
search spaces and cost functions that depend on the search space coordinates in highly nontrivial ways. We note that a different structure prediction technique, based on Monte Carlo (MC) simulations, has been introduced in the recent literature \cite{FilionMOPSD09}. For a thorough comparison between these two approaches, applied to patchy systems, we refer to Ref.~\cite{BianchiDFDK12}.

Working at constant pressure $P$, the lattice
parameters, the positions of the (up to eight \footnote[3]{For higher particle numbers, the computational cost of the optimizations rapidly increases. All the structures presented in this contribution can be realized with no more than four basis particles.}) particles within the
primitive cell and their orientations are optimized by minimizing the Gibbs
free energy, which reduces to the enthalpy at $T=0$.
To cope with the large number of parameters,
we use a phenotype implementation of
EA based optimization techniques~\cite{OganovG06,JPCM10}. 
In short, such a scheme combines evolutionary global optimization with local optimization steps. In the former, a set of candidate solutions (a ``population'' consisting of ``individuals''), each characterized by its search space coordinates and a fitness value (the latter being a function of its Gibbs free energy), is undergoing an evolutionary process; recombination and mutation operations are used in order to create new individuals, whose fitness values are expected to gradually improve. The local optimization is carried out by a quasi-Newton algorithm, namely L-BFGS-B \cite{ByrdLN95,ZhuBN97}.
The algorithm described
in Ref.~\cite{JPCM10} has been augmented in the following way:
(i) we make use of the angle-axis description for handling rotations of rigid bodies
first introduced for soft matter systems with anisotropic particles in Ref.~\cite{Wales05}. A detailed description and examples for successful application can be found in Refs.~\cite{ChakrabartiW09,FejerCW11}.
(ii) Since the time-consuming local optimization steps are
 independent from each other, they can be carried out
simultaneously on different processors. Further, we
drop the concept of generations in the algorithm and use an implementation of the pool based approach described in Ref.~\cite{BandowH06} instead.
(iii) Along the evolutionary process, bond order parameters~\cite{SteinhardtNR83, LechnerD08}
are used to distinguish between energetically equivalent, but structurally
different lattices in order to retain a diverse population by employing structural niches~\cite{Hartke99}.
The algorithm does not only keep track of the global minimum structure, but 
also of low-lying local minima, which might become dominant at finite $T$~\cite{Wales03}. 
As in any nondeterministic optimization approach there is no guarantee that the lowest identified 
minimum is indeed the global one, therefore, in a strict sense, we have to consider our structures putative groundstates~\cite{WalesS99}.
However, we affirmed our results by performing a large number of independent optimization cycles at each state point.

The structures suggested by the EA are replicated as starting configurations for NPT MC simulations, with typically about 500 particles in the simulation box.
The equations of state
of the fluid and solid phases are calculated using these simulations. For solids with non cubic symmetry the shape
of the simulation cell is allowed to change during the
simulations to obtain the equilibrium structure of the solid at 
each thermodynamic state.  Free energies of solids are calculated
using the Einstein molecule approach~\cite{VegaN07,review,JCP_130_244504_2009}, 
based on modifications of the Frenkel-Ladd
method~\cite{frenkel-ladd}.
The orientational degrees of freedom need special treatment, explained in further detail in the supplementary material$^{\dag}$. The free energy of the
liquid is obtained by thermodynamic integration from the ideal
gas. Once the free energy at a state point is known, free
energies at other states are calculated by
thermodynamic integration.  Coexistence points are obtained by
finding the temperature and pressure at which the chemical potential
of the two phases in coexistence are equal.  Once a coexistence point
is known, the whole coexistence line can be integrated by using the
Gibbs-Duhem method~\cite{kofkeJCP}.

\begin{figure*}[htbp]
\begin{center}
\includegraphics[width=7.0cm]{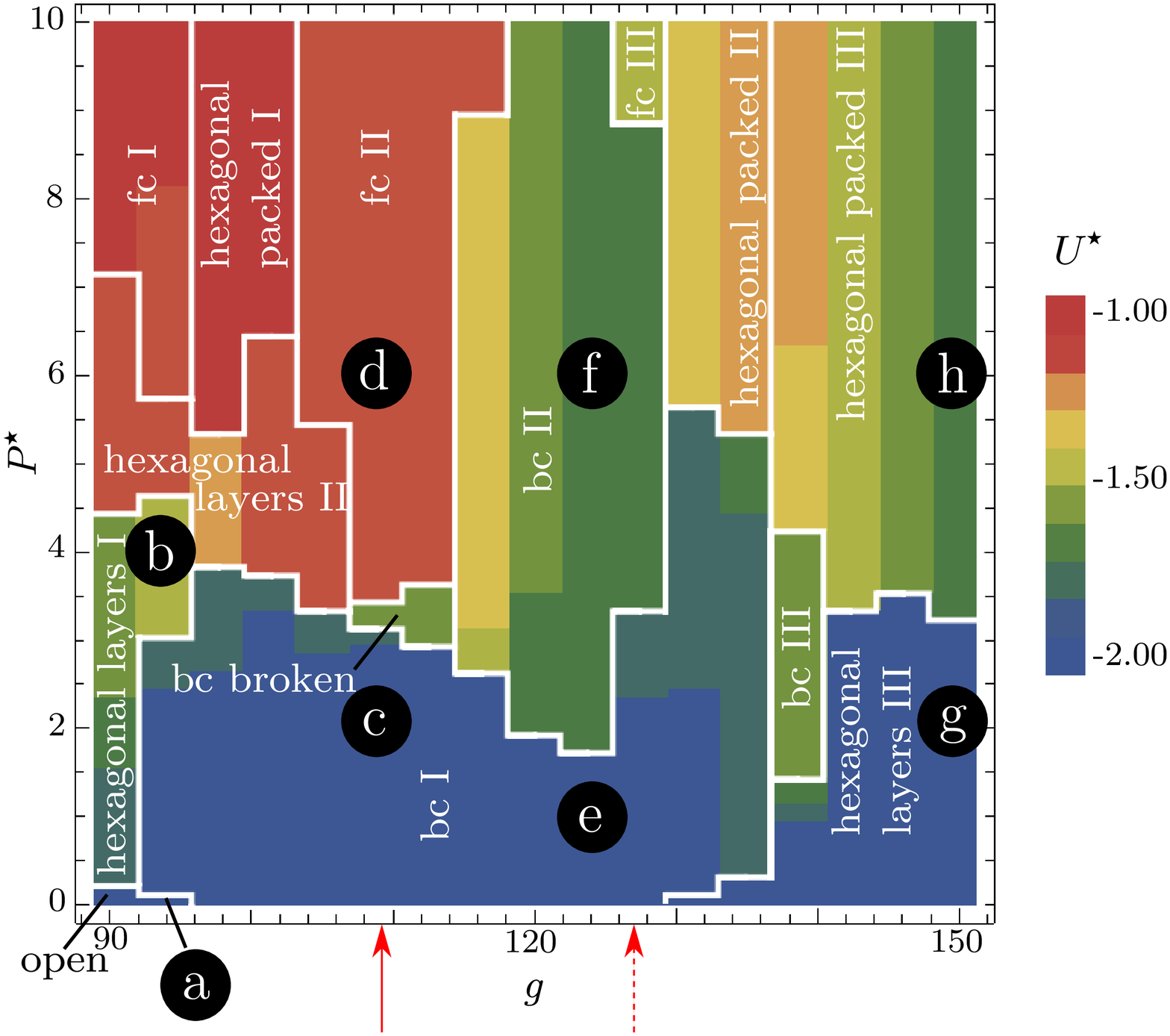}
\includegraphics[width=7.0cm]{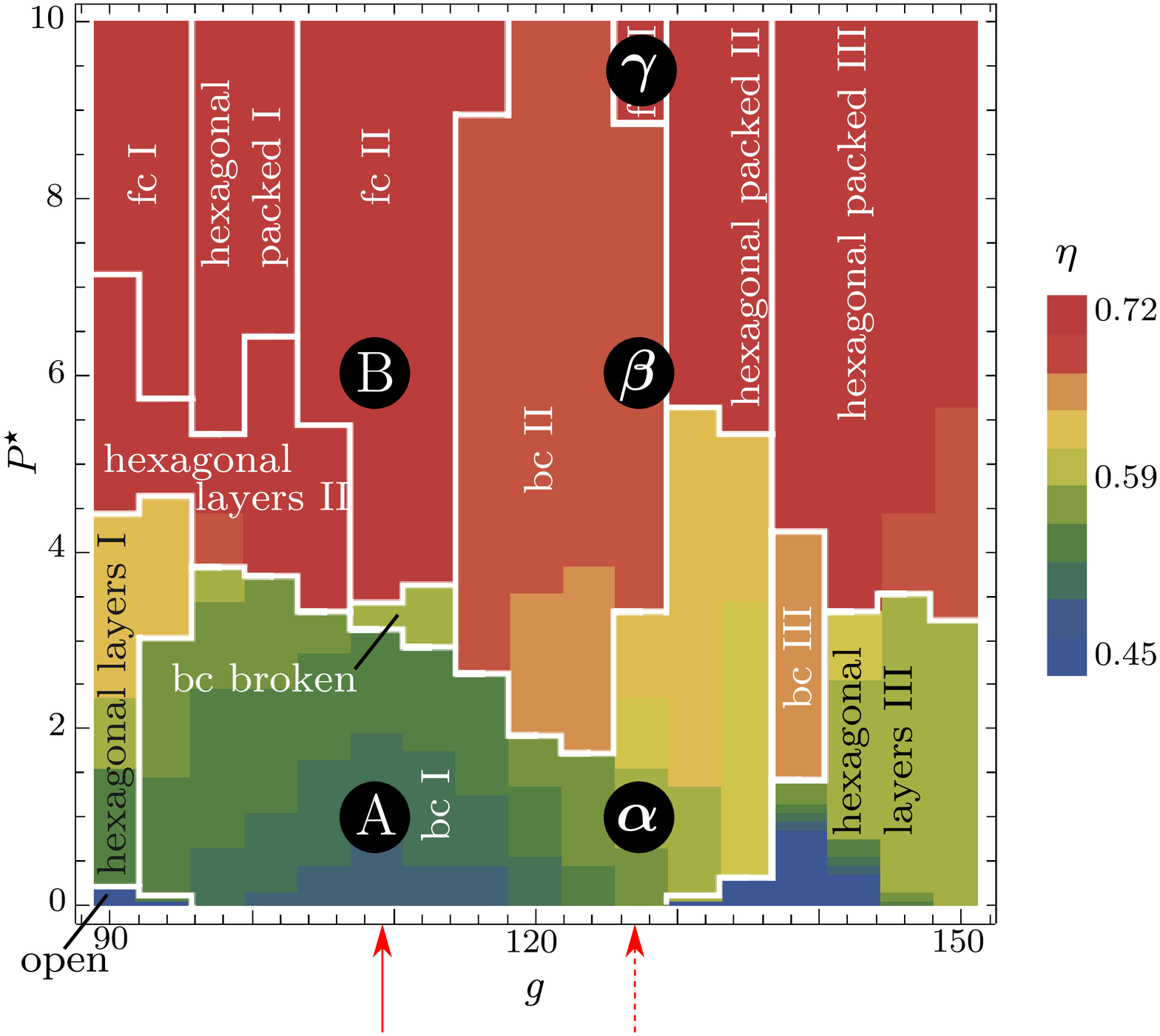}
\end{center}
  \caption{(Colour online) Contour plot of the reduced energy $U^{\star}$ (left) and packing fraction $\eta$ (right) as functions of the
    dimensionless pressure $P^{\star}$ and the geometric factor $g$; colour codes 
    are displayed (for $\eta$, the range is cut off at $\eta=0.45$, i.e., all values below 0.45 appear in the same colour). White boundaries indicate the
    limits of stability of the respective ordered structures on the
    underlying $(P^{\star}, g)$ grid. ``bc'' stands for body-centered lattices, ``fc'' for face-centered lattices. Solid arrows specify the
    $g$ value corresponding to a regular tetrahedral patch arrangement,
    dotted arrows correspond to $g=127.50$,
    for which we present results at finite $T$. Labels specify configurations referred to in the text.}
  \label{fig:landscapes}
\end{figure*}

We start the discussion of our results with the ordered equilibrium structures
at $T=0$\footnote[4]{Note that for many values of the geometry parameter, the patchy interactions lead to -- sometimes strong -- deviations from perfectly symmetrical lattices. Therefore we refer to the structures we identified as bcc-like, fcc-like or hcp-like to indicate such aberrations.}. The broad variety of identified lattice structures
is summarized
in Figure \ref{fig:landscapes}.
The topologies of the energy- and packing fraction-landscapes are the
result of a complex competition between two mechanisms:
packing (minimizing $V$) vs. bond saturation (minimizing the
energy $U^\star$). Accordingly, we can identify three regions in $(P^{\star}, g)$ space:

  (i) For pressure values up to $P^{\star} \simeq 4.00$ fully bonded
  structures (i.e., $U^{\star}
  \simeq -2.00$) dominate, which have rather small packing fractions,
  from $\eta \simeq 0.27$ for open structures to $\eta \simeq 0.64$ for distorted body-centered (bc)
  structures (compared to $\eta \simeq 0.74$ for close-packed spheres).
  The spectrum of identified lattices ranges from open, layered
  structures ($g \sim 90.00$) over bc lattices (which dominate
  over a broad $g$ range, i.e., $93.75 \lesssim g \lesssim
  135.00$) to layers of hexagonally arranged particles for $g \gtrsim
  135.00$.
 (ii) For $g \lesssim 120.00$ the transition from low- to
  high-pressure structures is characterized by an abrupt drop in
  the bond saturation, increasing from nearly full bonding (i.e., $U^{\star} \sim
  -2.00$) to a value of around $-1.00$. In this range of $g$, the location of the
  patches does not allow for the formation of
  strong bonds, thus pressure rather easily wins over bond saturation.

 (iii) In contrast, for $g \gtrsim 120.00$, $U^{\star}$ increases more
  smoothly with $P^\star$ and the identified ordered
  structures are the result of a delicate tradeoff between saturation
  and packing. In particular we emphasize that for selected $g$ values
  (i.e., $g \simeq 123.75$ and $g \simeq 150.00$) the respective patch
  decorations support both a high degree of bond saturation as well as
  the formation of high density lattices.
  As a consequence, the identified structures are able
  to persist even up to high pressure values (i.e., $P^{\star} \simeq 10.00$)
  while maintaining a relatively low binding energy (i.e.,
  $U^{\star} \simeq -1.65$).

The complex topology of the energy and packing fraction landscapes shown in
Figure \ref{fig:landscapes} is
elucidated in the following by a more detailed discussion of the
identified ordered equilibrium structures for four representative
$g$ values. Visual representations of the equilibrium structures at the state points specified in the left panel of Figure \ref{fig:landscapes} by labels ``a'' to ``h'' can be found in the supplementary material$^{\dag}$.

  For $g = 93.75$ a layered structure with full bond saturation
  ($U^{\star} \simeq -2.00$) is formed at very low pressure ($P^{\star} \lesssim 0.05$):
  each layer consists of
  fully bonded particles, forming a honeycomb lattice; the inter-layer
  bonding is realized via the patches located at the north poles (``a''). Upon compression,
  the system forms a hcp-like structure (i.e., AB stacking of hexagonally arranged layers) 
  which is characterized by rather weak \textit{intra}-layer ($U^{\star}
  \simeq -0.30$) and stronger \textit{inter}-layer ($U^{\star} \simeq -0.40$) bonds (``b'').

  For the case of a regular tetrahedral patch arrangement ($g \simeq
  109.47$), particles arrange -- as expected -- at low $P^\star$
  in a bcc-like lattice (``c''). 
  As discussed in Ref.~\cite{NoyaVDL10}, the bcc-like configuration can be seen as two 
  interpenetrating
diamond lattices, which only interact \textit{via} mutual repulsion at short distances. The optimized lattice at $T=0$ 
  shows another property: The two sublattices are slightly shifted against each other, i.e., 
  particles of sublattice A do not lie exactly in the center of the voids of sublattice 
  B and vice versa \footnote[5]{In Figure 2, central left panel, of the supplementary material$^{\dag}$, the yellow particles would lie exactly on top of the red particles in a perfect bcc lattice.}. By this rearrangement, optimal bond lengths can be retained at higher densities.
  At pressure values $3.20 \lesssim P^{\star} \lesssim 3.40$, a variant
  of the bcc-like structure, where one of the bonds is broken in
  order to increase the packing fraction by nine percent, is stable. 
  For larger $P^\star$, the transition into a compact fcc-like structure (with
  $U^{\star} \simeq -1.10$ and $\eta \simeq 0.71$, ``d'') takes place.
We note that hcp-like structures are also identified by the search algorithm as low-lying local minima at high pressure, but turn out to be thermodynamically unstable in this model. For a more detailed study, we refer to Ref.~\cite{lmcproceedings}.

  For $g = 123.75$, we
  encounter a lattice with the same bonding pattern as the aforementioned bcc/double 
  diamond lattice for low pressure values (``e'').  
  The
  ordered medium pressure phase is different from the one
  encountered for the regular tetrahedral patch decoration: particles
  self-assemble now in a rather closely packed bc-like lattice
  ($\eta \simeq 0.68$, ``f''). However, the particular patch decoration allows
  now for a considerably enhanced bond saturation: bonds are formed
  between second-nearest neighbours in the bcc-like lattice, i.e., along the edges of the bcc-like
  unit cell, leading to $U^{\star}
  \simeq -1.65$. This is the only configuration with $\eta < 0.70$ that is able
  to survive up to the highest pressure values investigated.
  
  Finally, for $g = 150.00$, the ordered
  low $P^\star$ phase can be viewed as a stacking of staggered,
  hexagonally ordered double layers: each double layer is formed by
  two congruent hexagonal particle arrangements, bonded to each
  other via the patch located at the north pole. The
  double layers themselves are connected via the three patches located
  at the basis; the
  bonds are \textit{fully} saturated (``g''). For the high pressure phase (i.e.,
  $P^{\star} \gtrsim 3.00$), the system forms an ABAB... stacking of hexagonally close-packed layers. 
  Like in the low pressure configuration, the particles are essentially unbonded within these layers, 
  but establish relatively strong bonds to the neighbouring layers (with $U^{\star} \simeq -1.65$, ``h'').
  In contrast to the low pressure structures, the particle axes are tilted against the normal vector of the layers, 
  thereby breaking one bond per particle and considerably decreasing the interlayer distance.

\begin{figure*}[htbp]
\begin{center}
\includegraphics[width=7.0cm]{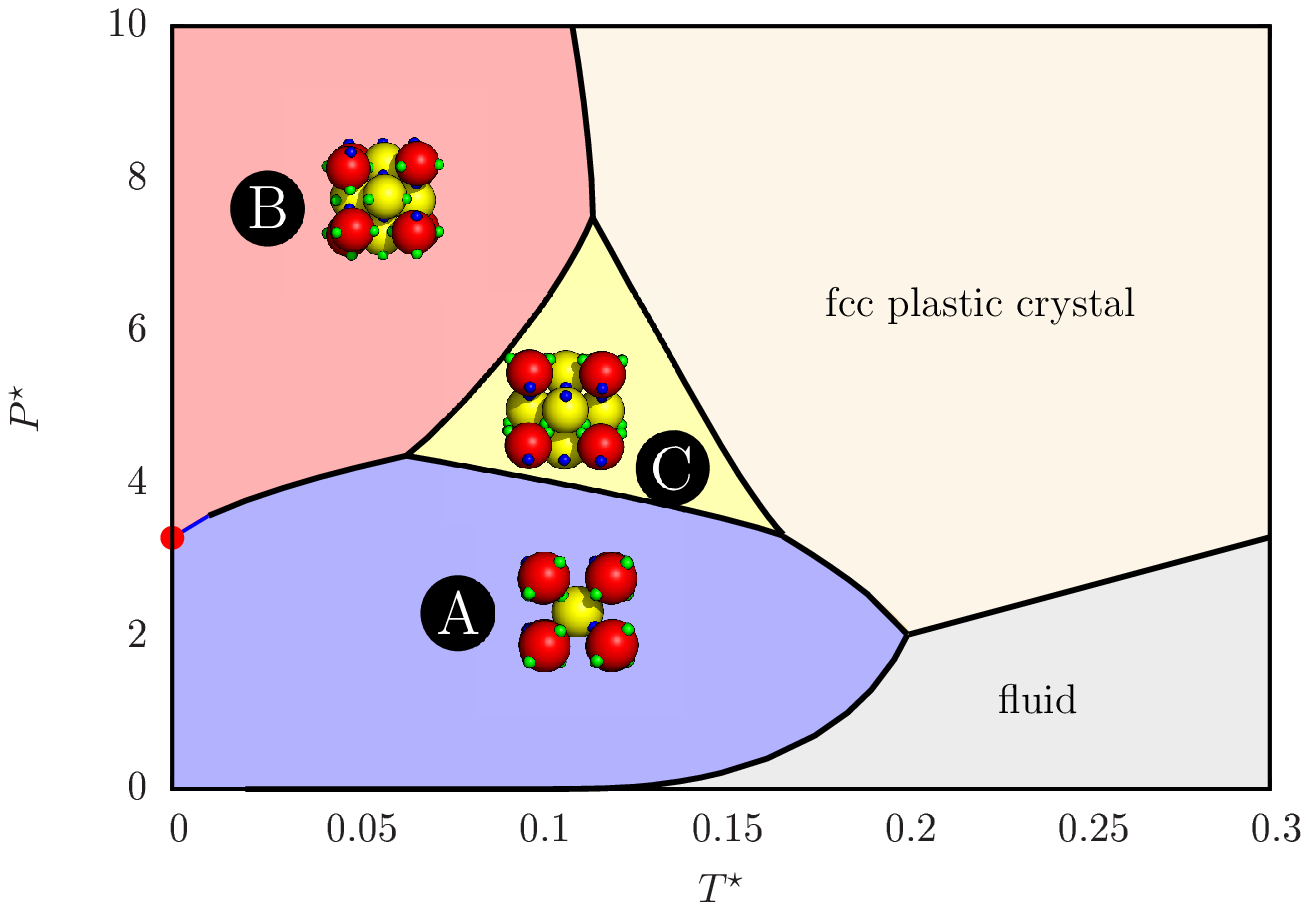}
\includegraphics[width=7.0cm]{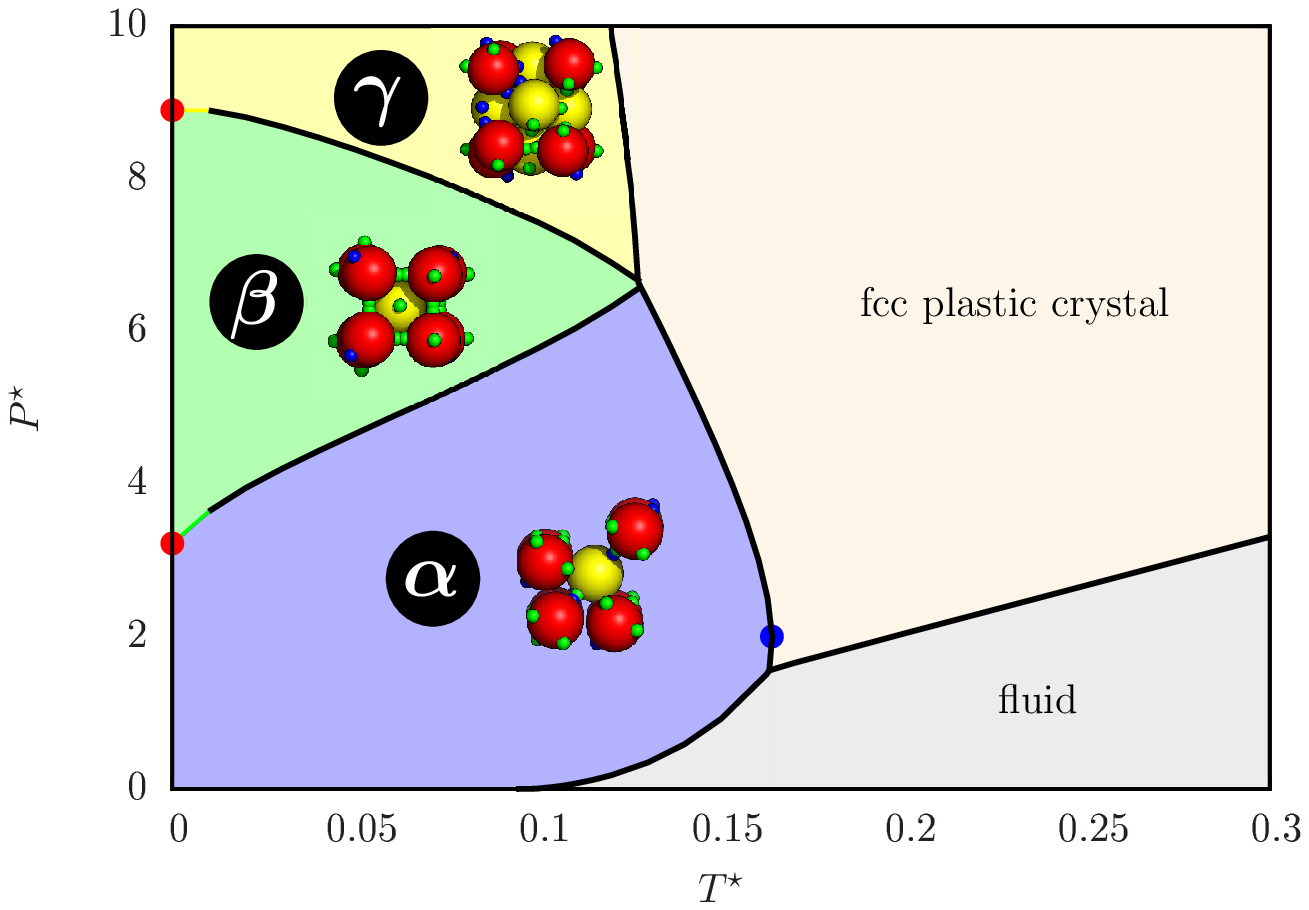}
\end{center}
  \caption{(Colour online) $(P^{\star}, T^{\star})$ phase diagram for $g \simeq 109.47$, corresponding to a regular
    tetrahedral arrangement of the patches (left) and $g \simeq 127.50$ (right).
    Coexistence lines are calculated down to $T^{\star}=0.01$.
    The dots on the y axes indicate the coexistence pressures at $T^{\star}=0$
    predicted by the optimization method (neglecting the bc-broken
    phase); the (blue) dot in the right panel indicates re-entrancy. Labels correspond to the ones in Figure \ref{fig:landscapes}, right panel.
    }
  \label{fig:pds}
\end{figure*}

With the candidate structures suggested by the optimization procedure
we can now proceed to the evaluation of the phase diagram at
finite temperature (i.e. $T^\star$  up to $0.3$). In addition to the energetic and packing contributions to the thermodynamic potential, entropic effects come into play and
configurations that correspond to \textit{local} enthalpy minima at $T=0$ might be stabilized at finite $T$. Thus we also
consider lattices that are identified by the optimization procedure as minima that differ in
enthalpy at most by twenty percent from the global minimum structure at some pressure value. A thorough investigation of
the competition between structures corresponding to local minima on the enthalpy landscape for a
specific geometry can be found in Ref.~\cite{lmcproceedings}. Here, we selected two patch decorations,
which we consider to be of particular interest. The
first one is specified by $g \simeq 109.47$, corresponding to the regular
tetrahedral patch arrangement. As the second one ($g = 127.50$), we chose 
a geometry for which three different stable configurations have been identified
at $T^{\star}=0$; among those is the aforementioned strongly bonded bc
structure, which can persist up to high pressure values.

When the particles are decorated via a regular tetrahedral patch
arrangement, the phase diagram shows (cf. Figure \ref{fig:pds}, left panel) that
two of the candidate structures predicted to be stable at $T^\star=0$ are able to 
survive over a relatively large
temperature range: the bcc-like/double diamond configuration at low
$P^\star$ (``A'' in Figure \ref{fig:pds}) and the fcc-like
configuration at high $P^\star$ (``B''). The broken
bcc-like/double diamond structure on the other hand plays only a minor role
in the temperature range considered: our calculations suggest that
it is only stable for $T^\star \lesssim 0.006$. On increasing
$T^\star$, an additional fcc-like phase (``C''), which is closely related to a
local minimum on the ($T^{\star}=0$) enthalpy landscape
\footnote[6]{More specifically, the fourth-lowest local minimum predicted by the optimization method is structurally and orientationally very similar and immediately relaxes to structure ``C'' in the MC simulations.}, becomes stable \cite{lmcproceedings}.
Its region of stability emerges out of a triple point ($T^\star=0.064, P^\star=4.39$) and rapidly
extends with increasing temperature at the cost of the two low
temperature phases. With further increasing temperature, the spatially and
orientationally ordered lattices transform via first order transitions
at low pressure into the liquid phase and at intermediate and high
pressure into an fcc structure with perfect symmetry, where the particles are
orientationally disordered (plastic crystal). 
At even higher temperatures
(not investigated here) the coexistence pressure of the plastic crystal and the fluid phase will gradually increase.

Diamond cubic and diamond hexagonal structures, which represent
local minima at very low $P^\star$ values, are never stable, but might be
stabilized by shifting the minimum of the interaction potential to
smaller inter-particle distances (cf. Ref.~\cite{NoyaVDL10}).
Comparing the present results
with the ones from Ref.~\cite{NoyaVDL10}, we note that the high density fcc-like
configuration (``B''), which has not been taken into account for the
previous calculations, makes the phase diagram considerably more complex.

The phase diagram of the second system investigated (cf. Figure \ref{fig:pds}, right panel)
displays some additional interesting features. With increasing
$T^\star$, the predicted  low- and high-pressure phases (bc-like, ``$\alpha$'' and fcc-like,
``$\gamma$'', respectively in Figure \ref{fig:pds}) extend their region of stability at the
cost of the region of stability of the high density bcc-like phase
(``$\beta$''). This region terminates in two very close triple points (``$\alpha$''-``$\beta$''-``$\gamma$'' and ``$\alpha$''-``$\gamma$''-fcc plastic crystal). At low $P^\star$, the ``$\alpha$'' phase transforms into the fluid phase, while at intermediate and high $P^\star$, the ``$\alpha$''and the ``$\gamma$'' phases transform via first order phase transitions into the fcc plastic crystal phase. We also observe a re-entrant scenario, which occurs along the coexistence line between the plastic
crystal and the ``$\alpha$'' phase. In a narrow temperature range (i.e., for
$0.163 \lesssim T^\star \lesssim 0.164$) the orientational disorder is first
replaced by orientational order and is eventually re-established upon
lowering pressure. 
The origin of this behaviour is the higher compressibility of
the fcc plastic crystal compared to the bc-like ``$\alpha$'' solid. The latter
structure is stabilized by its low energy, which is only obtained when the bond length is close to the LJ minimum.

The high degree of internal consistency between the
two methodological approaches combined in this
contribution is corroborated by the fact that the coexistence lines
evaluated via simulations, tend towards the coexistence pressure value
predicted by the optimization procedure for $T^\star \rightarrow 0$
(indicated by dots on the $y$-axes in Figure \ref{fig:pds}).

The detailed structural analysis and the phase diagrams presented in
this contribution have provided quantitative
evidence of the complex phase behaviour and the broad variety of
ordered equilibrium structures into which patchy particles are able to
self-assemble. These highly complex configurations are the result of an
intricate competition between bond formation, packing and entropy. The
possibility to discriminate between different structural particle
arrangements that are separated only by minute energy differences in
combination with its reliability make our conceptual approach highly
suitable for future structural and thermodynamic investigations of
patchy particle systems.

Financial support by the Austrian Science Foundation (FWF) under
Project Nos. W004, P23910-N16 and M1170-N16 are gratefully acknowledged. E.G.N. acknowledges support from Grants Nos. FIS2010-15502 from the Direcci\'{o}n General de Investigaci\'{o}n and S2009/ESP-1691 (program MODELICO) from the Comunidad Aut\'{o}noma de Madrid.

\balance
\footnotesize{
\providecommand*{\mcitethebibliography}{\thebibliography}
\csname @ifundefined\endcsname{endmcitethebibliography}
{\let\endmcitethebibliography\endthebibliography}{}

}

\onecolumn

\noindent\LARGE{\textbf{Supplementary information}}
\vspace{0.6cm}

\noindent\large{\textbf{}}\vspace{0.5cm}

\section*{Orientational reference Hamiltonian}

When calculating free energies of systems consisting of anisotropically interacting particles \textit{via} the Frenkel-Ladd method, the reference Hamiltonian includes an orientational term. The contribution to the entropy stemming from the orientational degrees of freedom
can be accounted for by integrating to a reference Einstein crystal in which the particles
are coupled to an orientational field that tends to align the particles with the same
orientation as in the solid structure under study.
It is usually convenient that the orientational field exhibits the same symmetry as the particles\cite{review}.
Therefore we used two different orientational fields for each of the geometries of
the particles for which the phase diagram at finite T was computed.
For the perfect tetrahedral case (with $T_d$ symmetry), 
the orientation of each particle in the reference structure can specified 
by two unitary vector ${\bf a_0}$ and ${\bf b_0}$ that are colinear with any
two specified patches, and the orientational field is defined as:\cite{NoyaVDL10,review}
\begin{equation}
U_{\rm orient} = \sum_{i=1}^N \lambda_o \left[ \sin^2 
\left( \Psi_{a,i} \right) + \sin^2 \left(\Psi_{b,i}\right) \right]
\label{field_orient1}
\end{equation}
where $\lambda_o$ is the coupling parameter, which has units of energy, and
$\Psi_{a,i}$ ($\Psi_{b,i}$) is the angle formed by the closest patch in the instantaneous
orientation of molecule $i$ and the vector ${\bf a_0}$ ($\Psi_{b,i}$) in the reference structure.
The particles with $g\approx$ 127.50 have $C_{3v}$ symmetry. In this case the 
orientation of the particles is defined by a vector colinear to the patch at the "north pole" (${\bf a_0}$) 
and the other is colinear to one of the other three patches (${\bf b_0}$).
The orientational field is the defined as:
\begin{equation}
U_{\rm orient} = \sum_{i=1}^N \lambda_o \left[ 
\left( \frac{\Psi_{a,i}}{\pi} \right)^2 + \sin^2 \left(\Psi_{b,i}\right) \right]
\label{field_orient2}
\end{equation}


\large

\clearpage

\section*{Visual representations of selected ordered equilibrium structures at $T=0$}

In Figures 1 to 4, we display the ordered equilibrium structures identified at $T=0$ for those systems that are specified in the panels of Figure 2 of the main article by labels ``a'' to ``h''. For the convenience of the reader, we repeat here the descriptions of these strucutres provided in the main article.
\begin{itemize}
\item
``a - open'': layered structure with full bond saturation;
  each layer consists of
  fully bonded particles, forming a honeycomb lattice; the inter-layer
  bonding is realized via the patches located at the north poles
\item
``b - hexagonal layers I'': hcp-like structure (i.e., AB stacking of hexagonally arranged layers) 
  which is characterized by rather weak \textit{intra}-layer and stronger \textit{inter}-layer bonds
\item
``c - bc I'': bcc-like configuration; can be seen as two 
  interpenetrating diamond lattices, which only interact \textit{via} mutual repulsion at short distances; the two sublattices are slightly shifted against each other, i.e., particles of sublattice A do not lie exactly in the center of the voids of sublattice B and vice versa, as can be seen in the central left panel of figure \ref{fig:g_109}: the yellow particles would lie exactly on top of the red particles in a perfect bcc lattice
\item
``d - fc II'': compact, almost close-packed fcc-like structure
\item
``e - bc I'': lattice with the same bonding pattern as structure ``c'' (bcc/double diamond), but slightly different arrangement of the colloidal particles
\item
``f - bc II'': rather closely packed bc-like lattice;
  the particular patch decoration allows
  for a high bond saturation: bonds are formed
  between second-nearest neighbours in the bcc-like lattice, i.e., along the edges of the bcc-like unit cell
\item
``g - hexagonal layers III'': stacking of staggered,
  hexagonally ordered double layers: each double layer is formed by
  two congruent hexagonal particle arrangements, bonded to each
  other via the patch located at the north pole; the
  double layers themselves are connected via the three patches located
  at the basis; all bonds are fully saturated 
\item
``h -hexagonal packed III'': ABAB... stacking of hexagonally close-packed layers; particles are essentially unbonded within these layers, 
  but establish relatively strong bonds to the neighbouring layers 

\end{itemize}

\clearpage

\begin{figure}[htbp]
\begin{center}
\begin{tabular}{c c}
\parbox[c]{1.1cm}{\includegraphics[width=1.0cm]{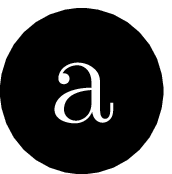}} &
\parbox[c]{6.1cm}{\includegraphics[width=6.0cm]{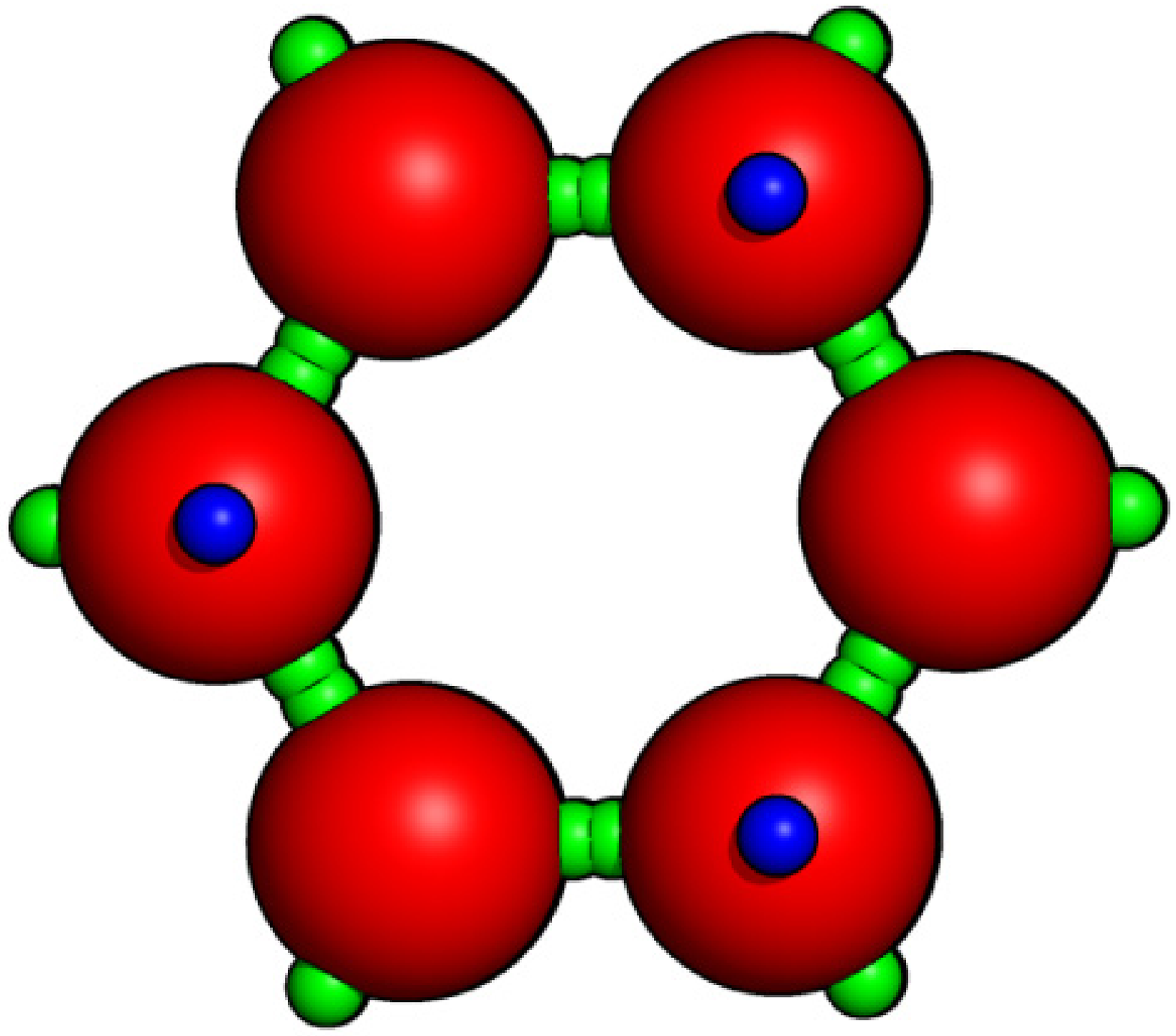}} 
\parbox[c]{6.1cm}{\includegraphics[width=6.0cm]{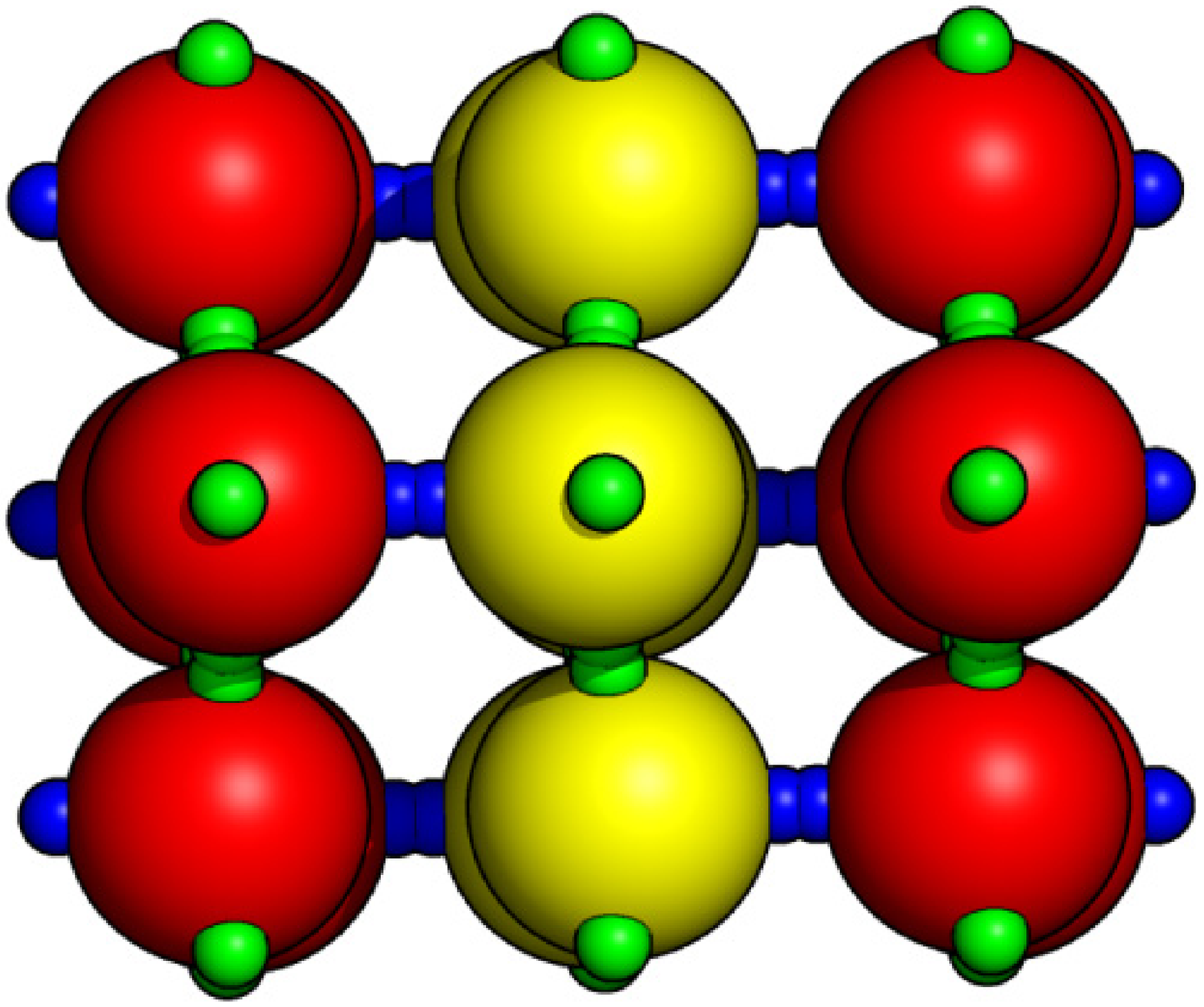}} \\
\hline
\parbox[c]{1.1cm}{\includegraphics[width=1.0cm]{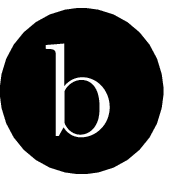}} &
\parbox[c]{6.1cm}{\includegraphics[width=6.0cm]{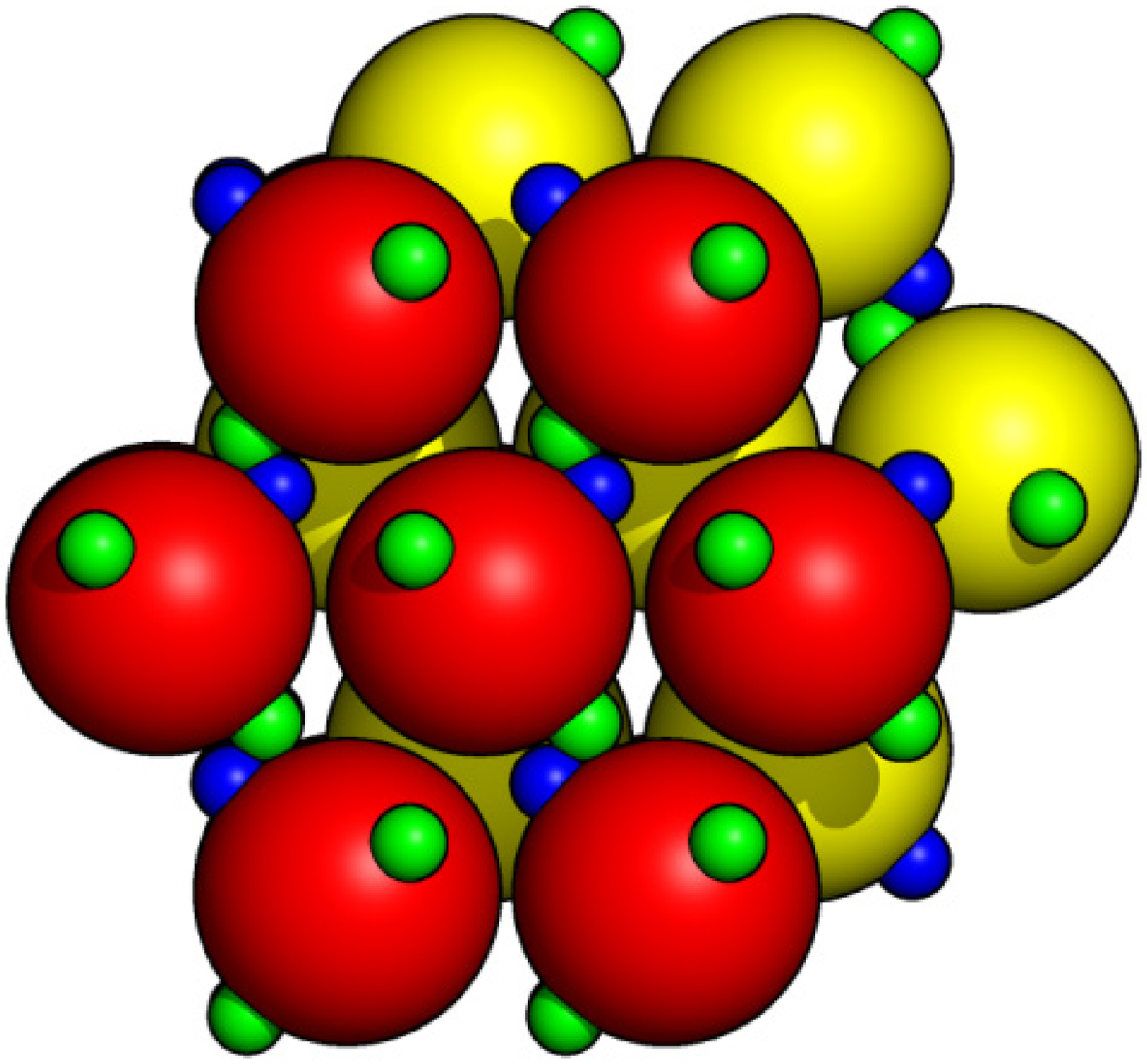}} 
\parbox[c]{6.1cm}{\includegraphics[width=6.0cm]{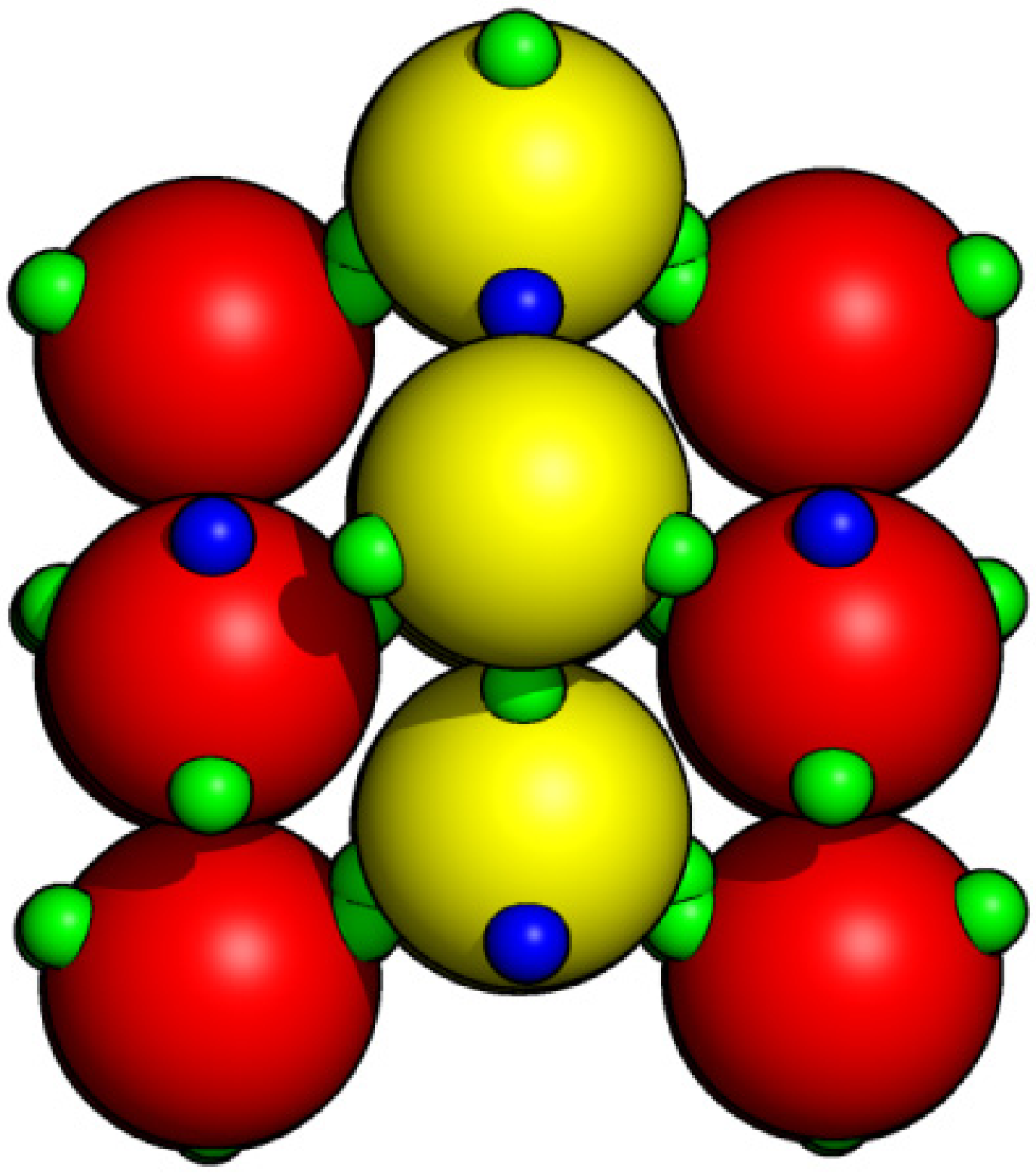}}
\end{tabular}
\end{center}
  \caption{\large Perpendicular views of the low-pressure (top) and the
    high-pressure (bottom) ordered equilibrium structure obtained for
    $g = 93.75$ (labels ``a - open'' and ``b - hexagonal layers I'' in Figure
    2 of the main article). The colour code for the blue and
    green patches has been specified in Figure 1 of the main article, the
    colours red and yellow for the patchy particles have been
    introduced for convenience$^\ast$.}
  \label{fig:g_90}
\end{figure}

\footnotetext{\normalsize $\ast$ (i) For layered honeycomb lattices and structures consisting of hexagonal layers, particles located in different layers are coloured in an alternating pattern. (ii) For bcc-like lattices, particles located on the vertices of the cube appear in red, while the particles at the center of the cube are colour yellow. (iii) In the double diamond picture, particles belonging to different non-interacting diamond sublattices appear in different colours. (iv) For fcc-like structures, the particles located at the vertices of the cube and the particles at the centers of the faces of the cube are coloured red and yellow, respectively.}

\begin{figure}[htbp]
\begin{center}
\begin{tabular}{c c}
\parbox[c]{1.1cm}{} & 
\parbox[c]{6.1cm}{\includegraphics[width=6.0cm]{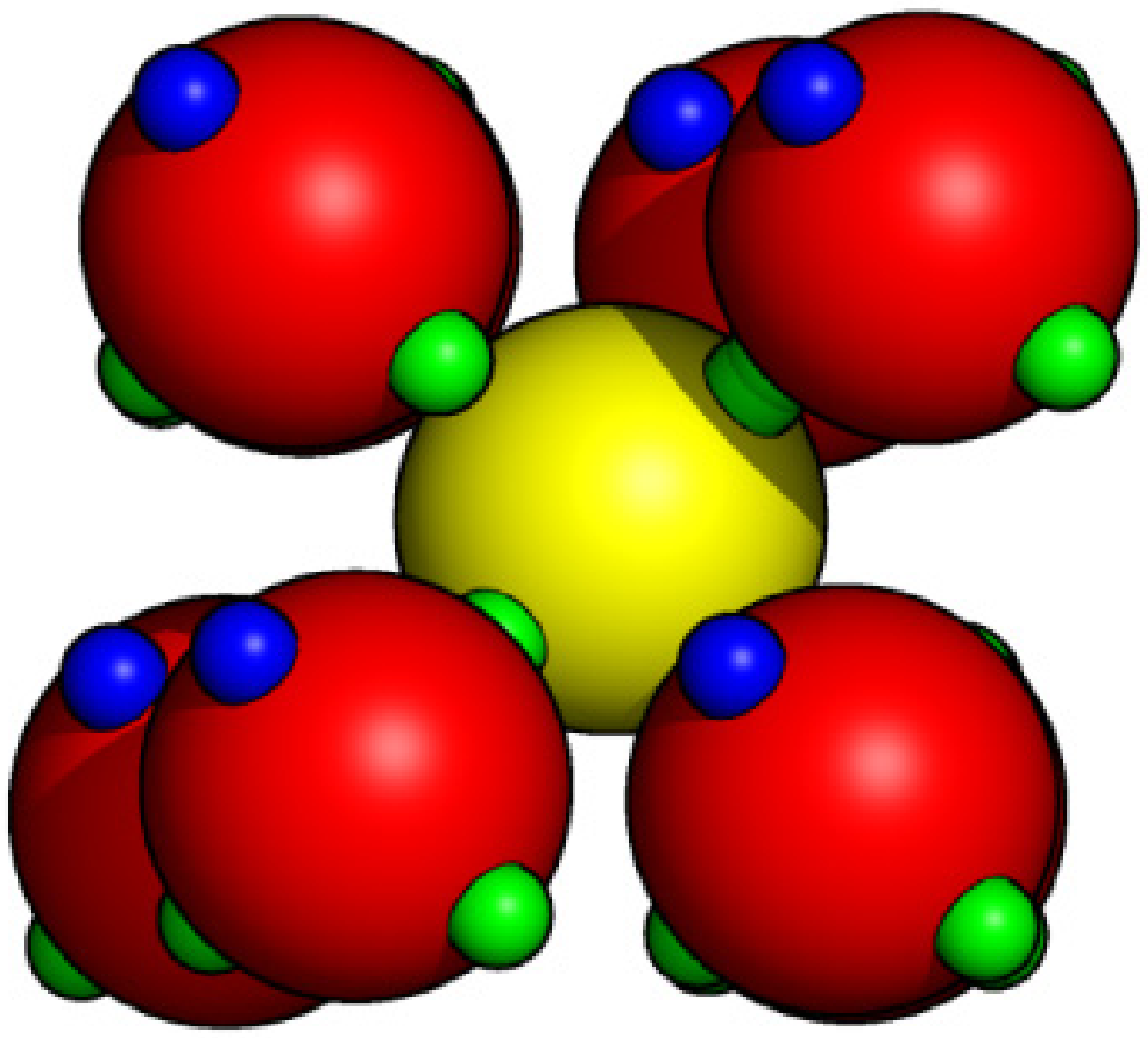}} 
\parbox[c]{6.1cm}{\includegraphics[width=6.0cm]{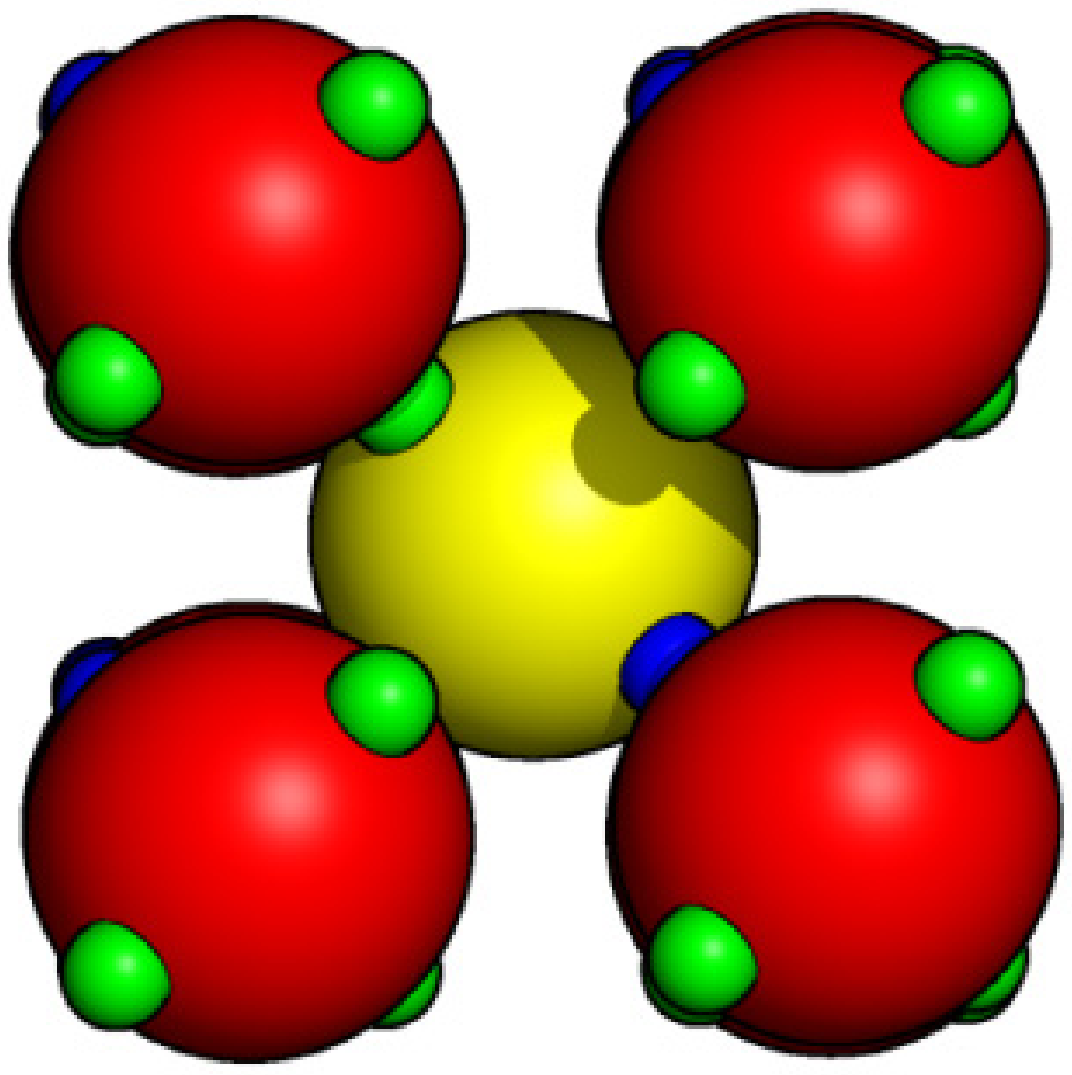}} \\
\parbox[c]{1.1cm}{\includegraphics[width=1.0cm]{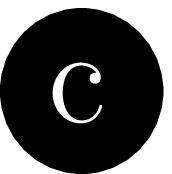}} & \\
\parbox[c]{1.1cm}{} & 
\parbox[c]{6.1cm}{\includegraphics[width=6.0cm]{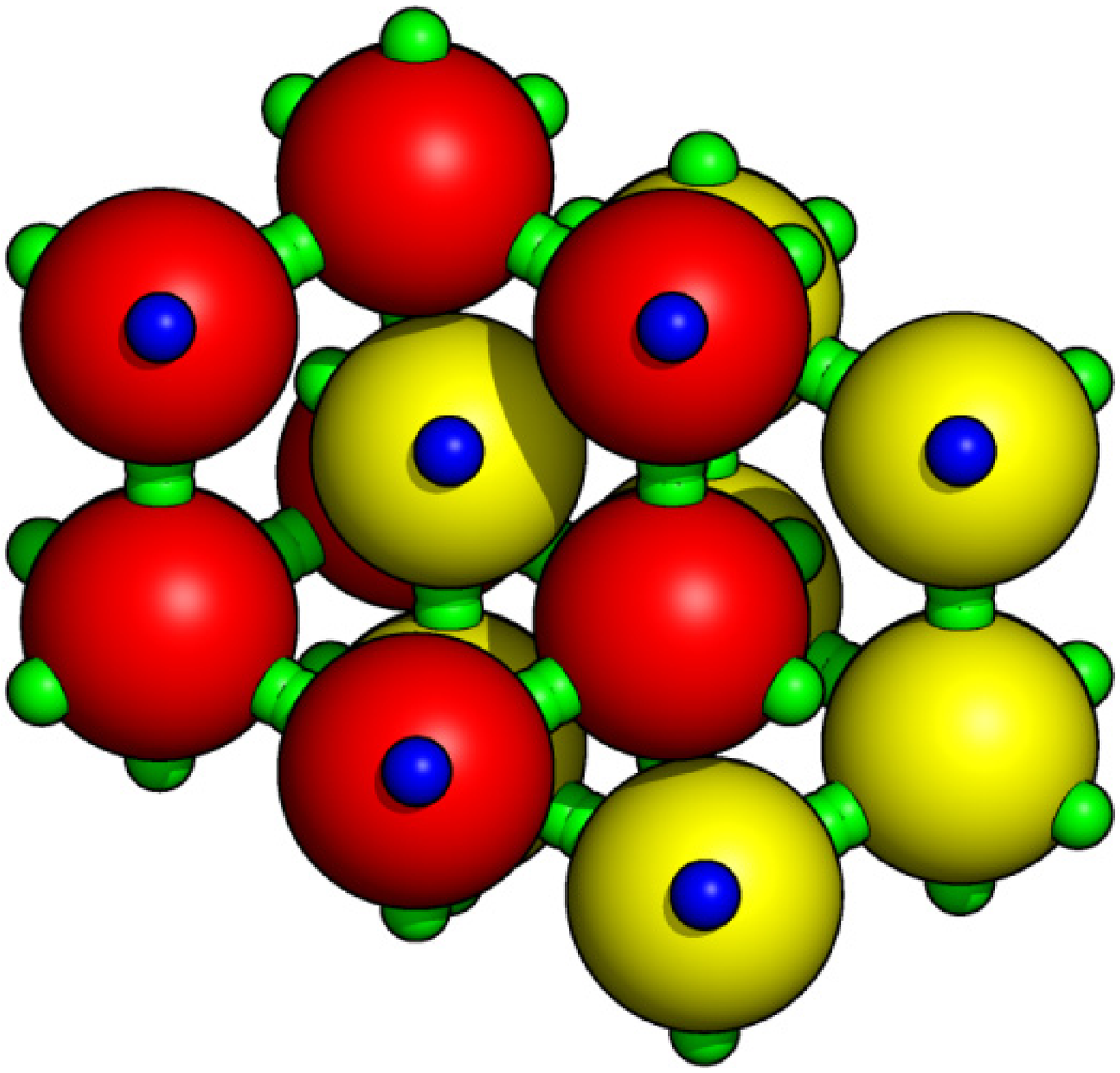}} 
\parbox[c]{6.1cm}{\includegraphics[width=6.0cm]{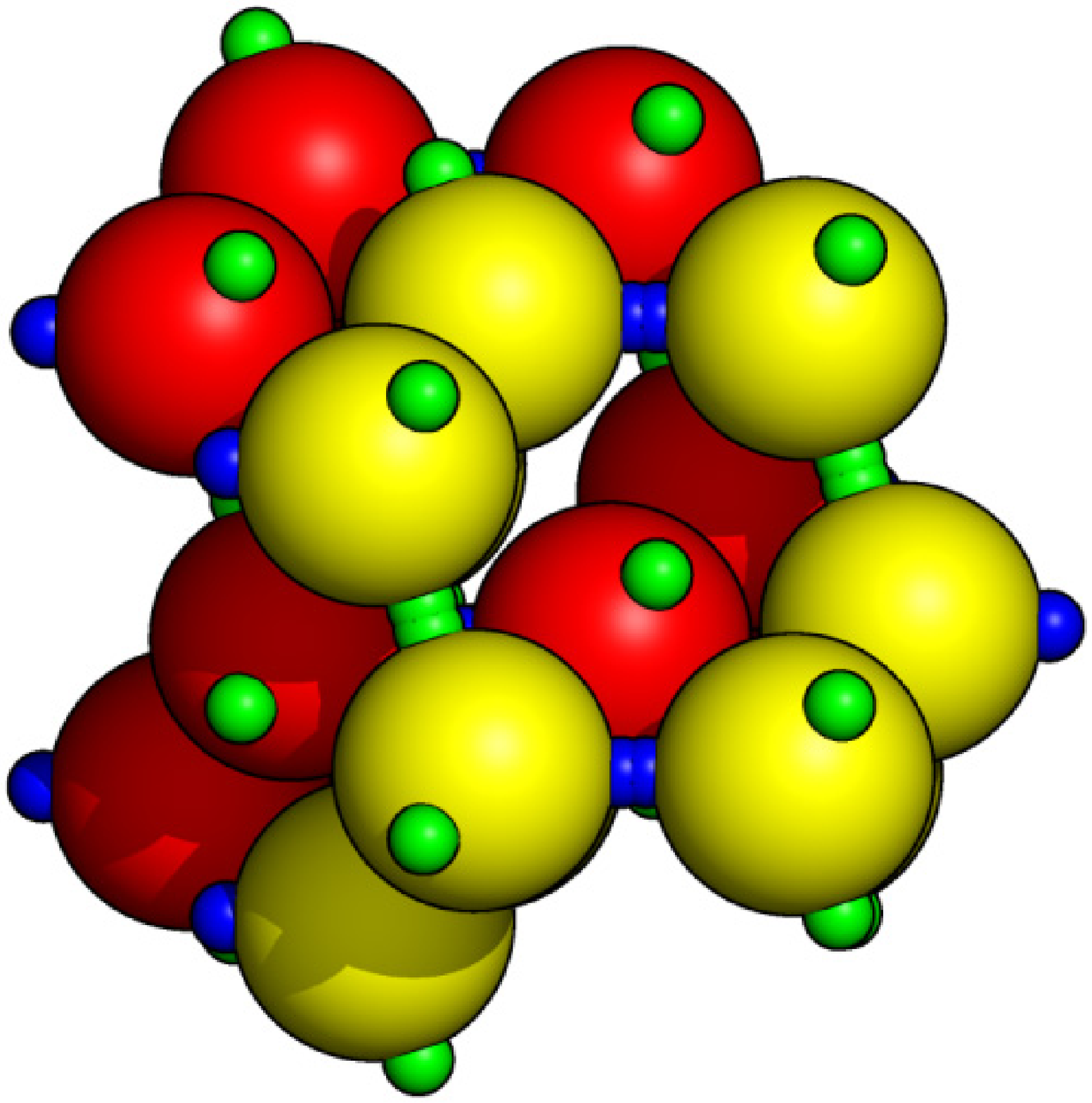}} \\
\hline
\parbox[c]{1.1cm}{\includegraphics[width=1.0cm]{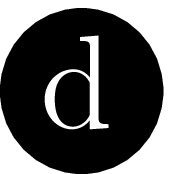}} & 
\parbox[c]{6.1cm}{\includegraphics[width=6.0cm]{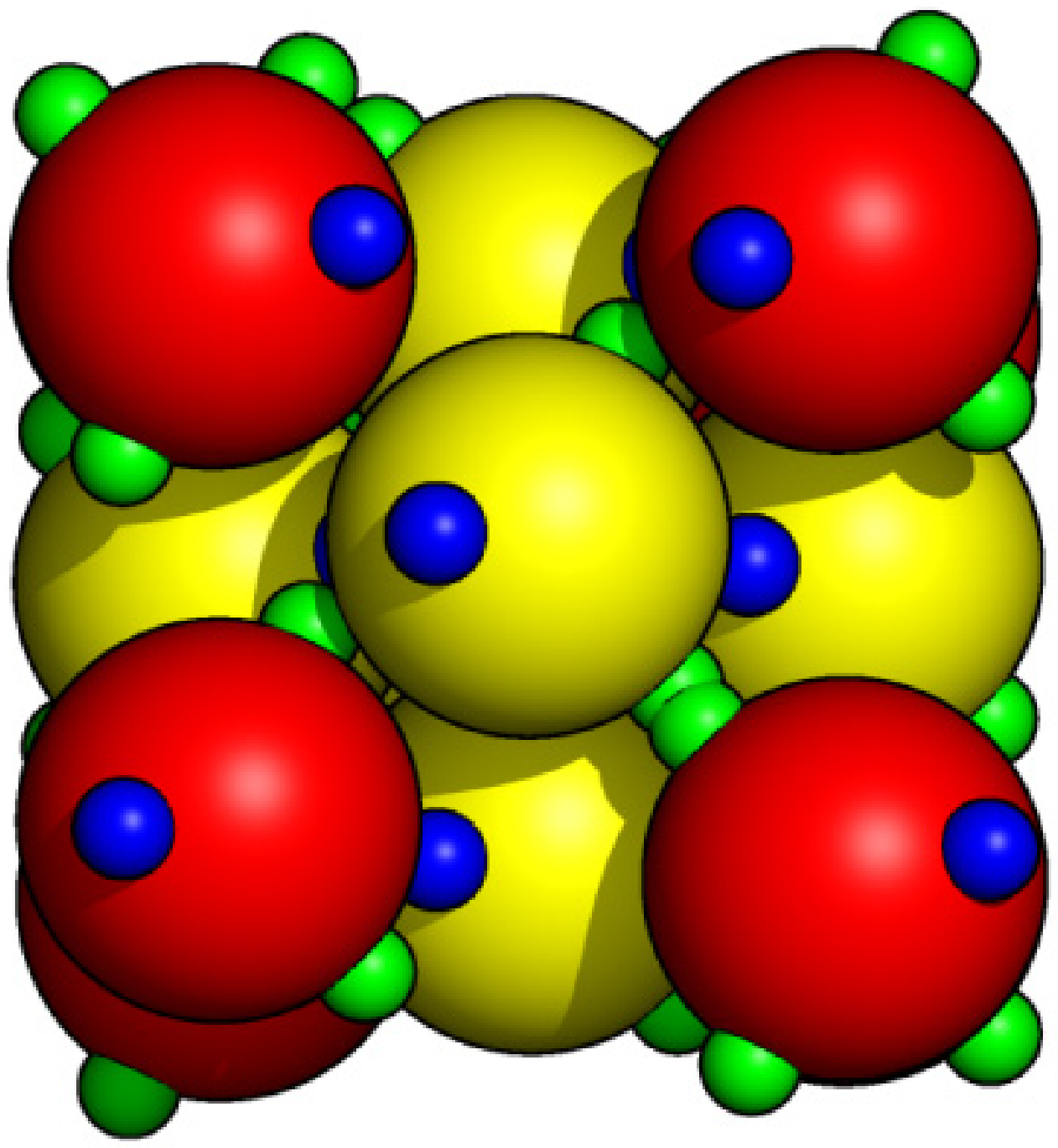}} 
\parbox[c]{6.1cm}{\includegraphics[width=6.0cm]{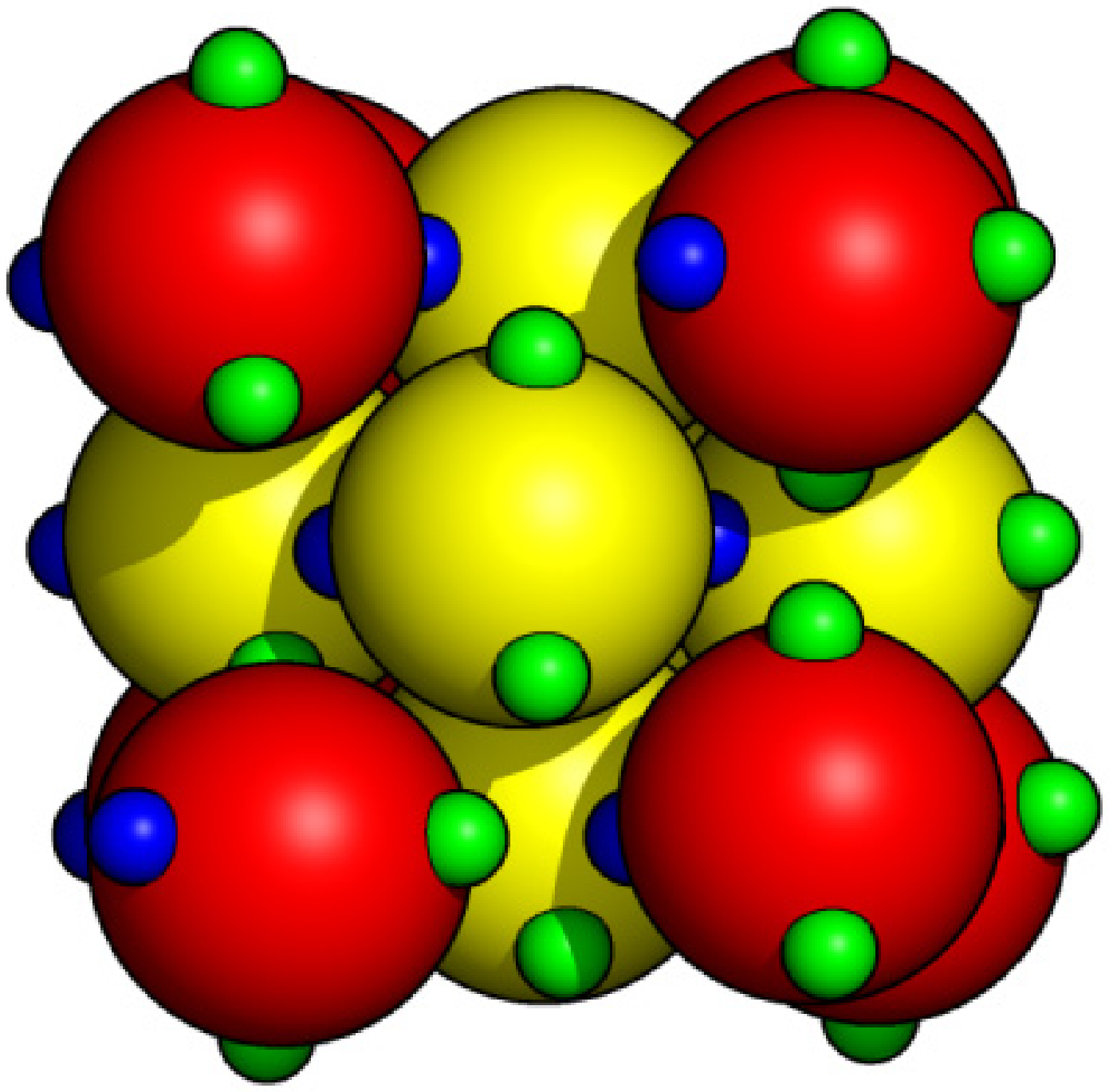}} 
\end{tabular}
\end{center}
  \caption{\large Perpendicular views of the low-pressure ordered equilibrium structure in the body-centered picture (top) and the double-diamond picture (center) as well as the  
    high-pressure structure (bottom) obtained for
    $g \simeq 109.47$ (labels ``c - bc I'' and ``d - fc II'' in Figure
    2 of the main article), corresponding to a regular
    tetrahedral arrangement of the patches. The colour code for the
    blue and green patches has been specified in Figure
    1 of the main article, the colours red and yellow for the patchy
    particles have been introduced for convenience$^\ast$.}
  \label{fig:g_109}
\end{figure}

\begin{figure}[htbp]
\begin{center}
\begin{tabular}{c c}
\parbox[c]{1.1cm}{\includegraphics[width=1.0cm]{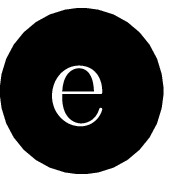}} &
\parbox[c]{6.1cm}{\includegraphics[width=6.0cm]{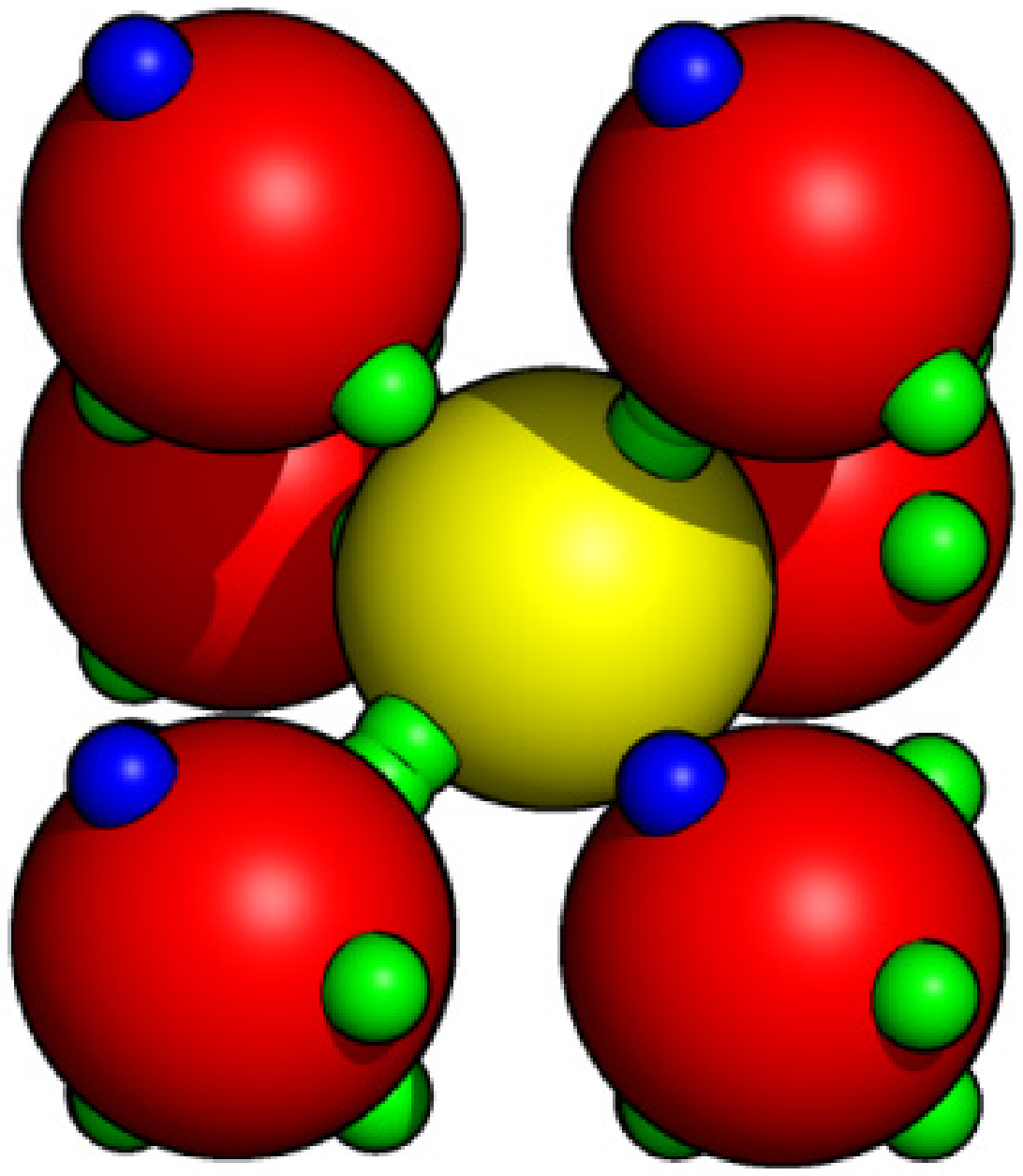}} 
\parbox[c]{6.1cm}{\includegraphics[width=6.0cm]{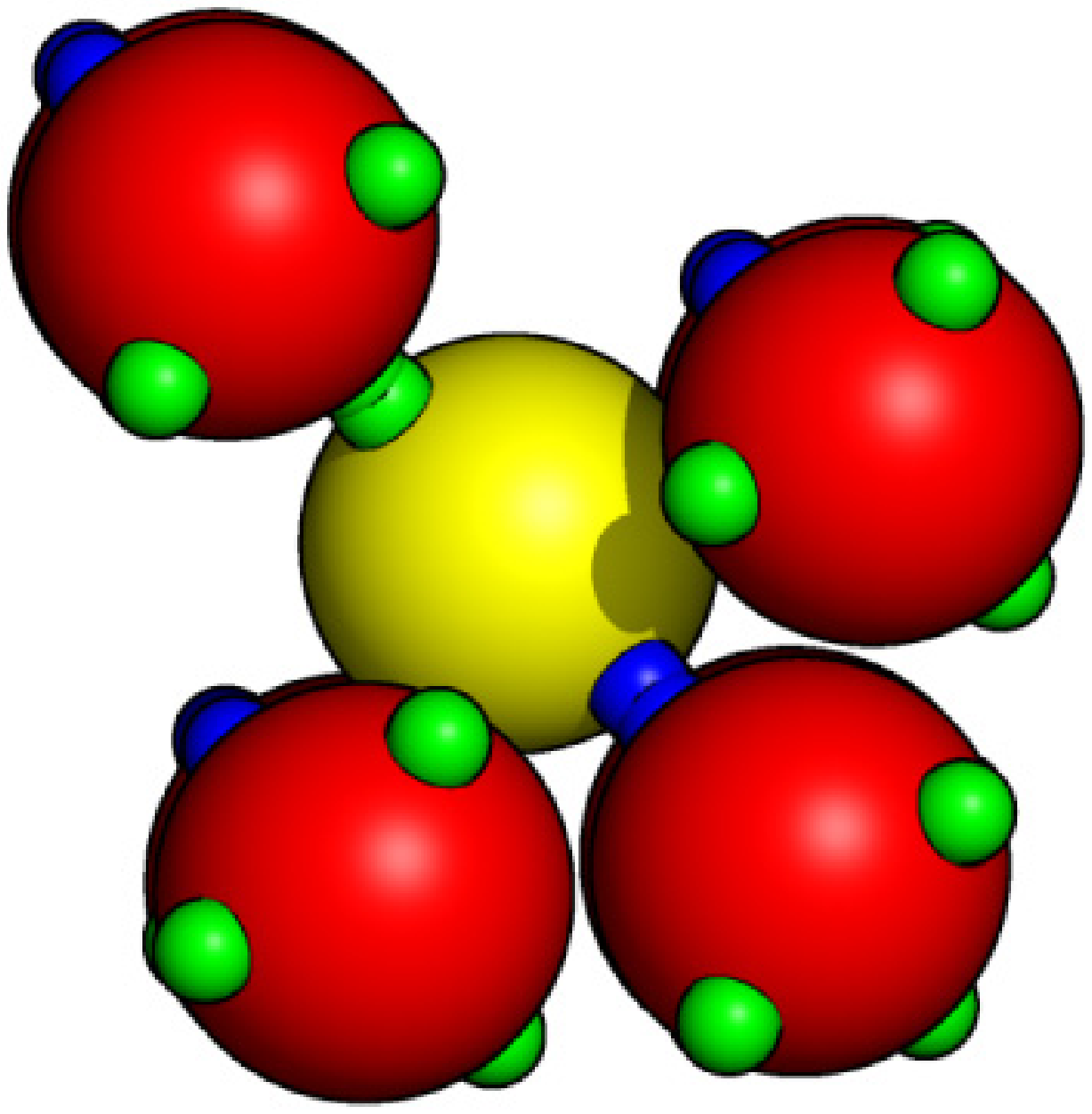}} \\
\hline
\parbox[c]{1.1cm}{\includegraphics[width=1.0cm]{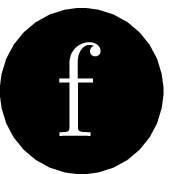}} &
\parbox[c]{6.1cm}{\includegraphics[width=6.0cm]{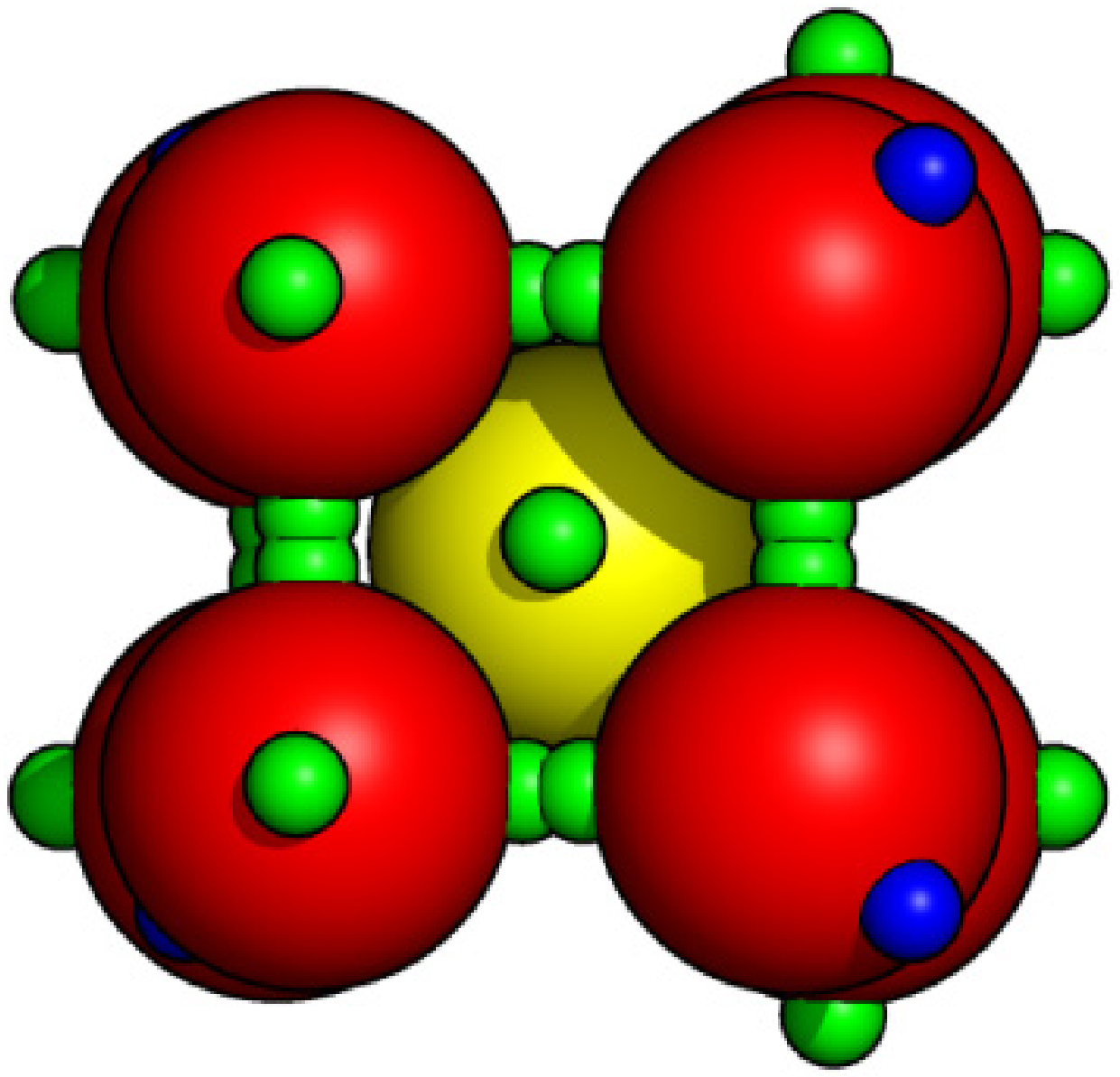}} 
\parbox[c]{6.1cm}{\includegraphics[width=6.0cm]{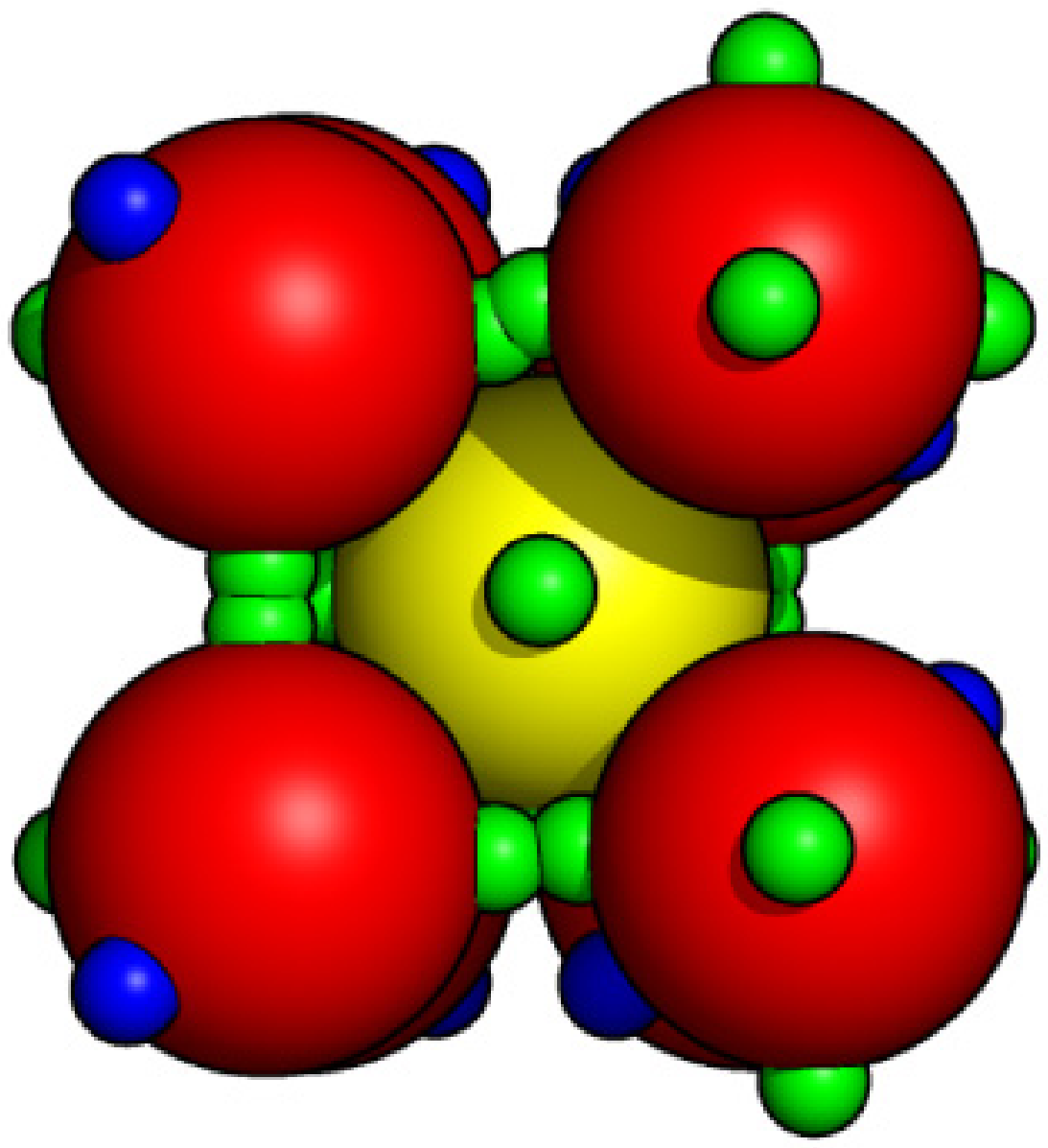}}
\end{tabular}
\end{center}
  \caption{\large Perpendicular views of the low-pressure (top) and the
    high-pressure (bottom) ordered equilibrium structure obtained for
    $g = 123.75$ (labels ``e - bc I'' and ``f - bc II'' in Figure
    2 of the main article). The colour code for the blue and
    green patches has been specified in Figure 1 of the main article, the
    colours red and yellow for the patchy particles have been
    introduced for convenience$^\ast$.}
  \label{fig:g_124}
\end{figure}

\begin{figure}[htbp]
\begin{center}
\begin{tabular}{c c}
\parbox[c]{1.1cm}{\includegraphics[width=1.0cm]{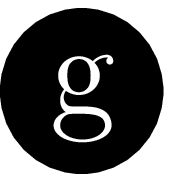}} &
\parbox[c]{6.1cm}{\includegraphics[width=6.0cm]{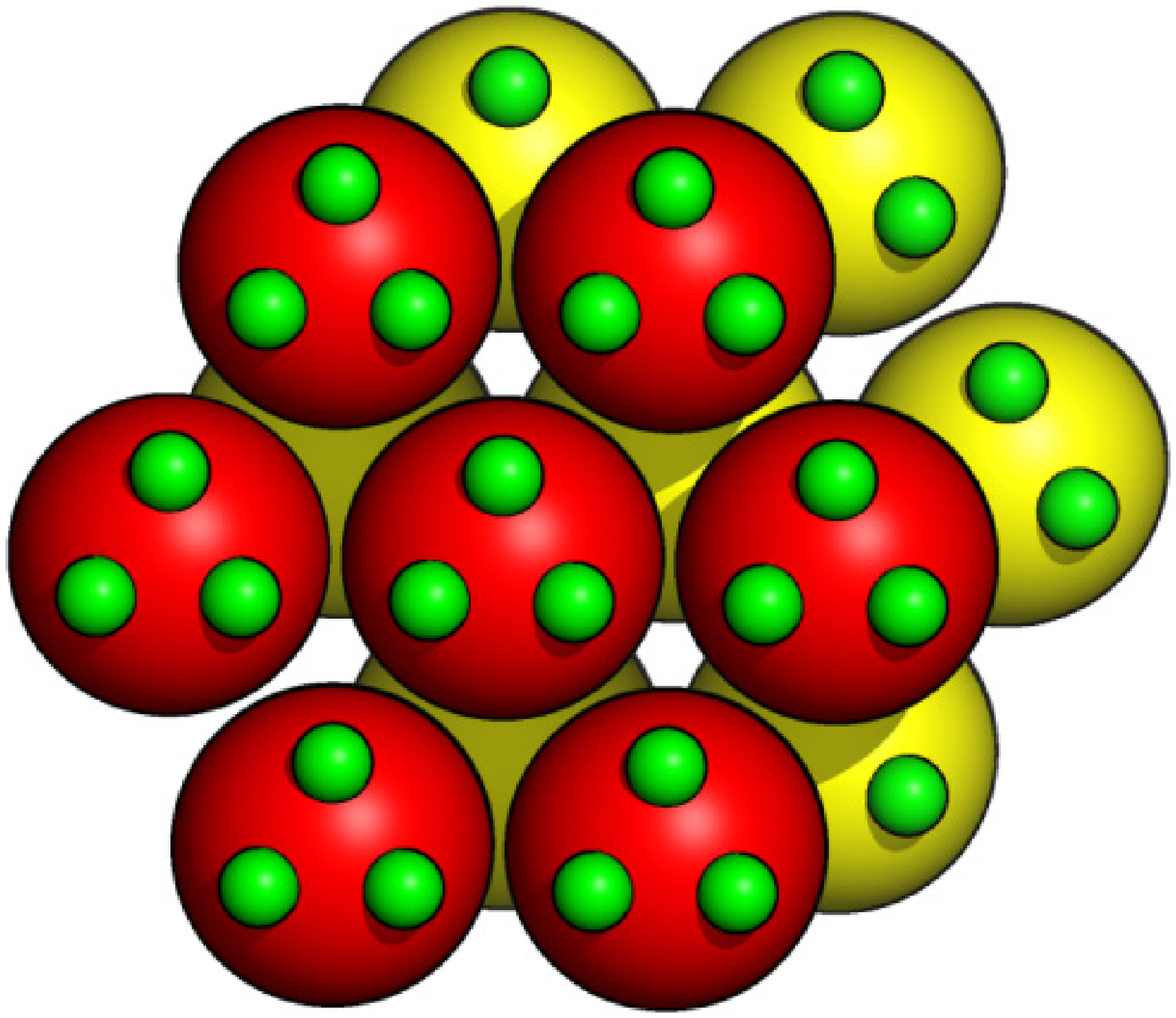}} 
\parbox[c]{6.1cm}{\includegraphics[width=6.0cm]{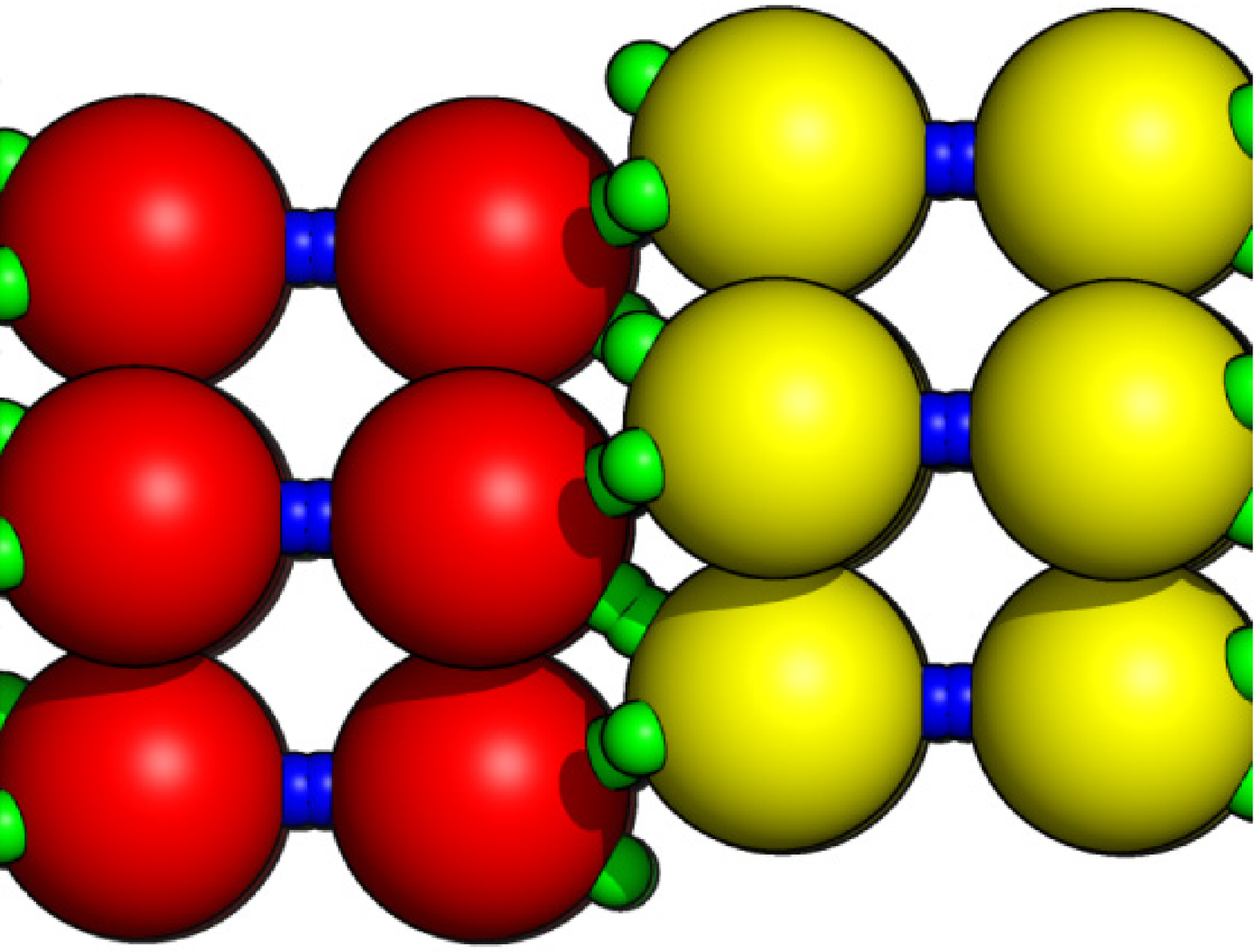}} \\
\hline
\parbox[c]{1.1cm}{\includegraphics[width=1.0cm]{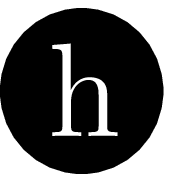}} &
\parbox[c]{6.1cm}{\includegraphics[width=6.0cm]{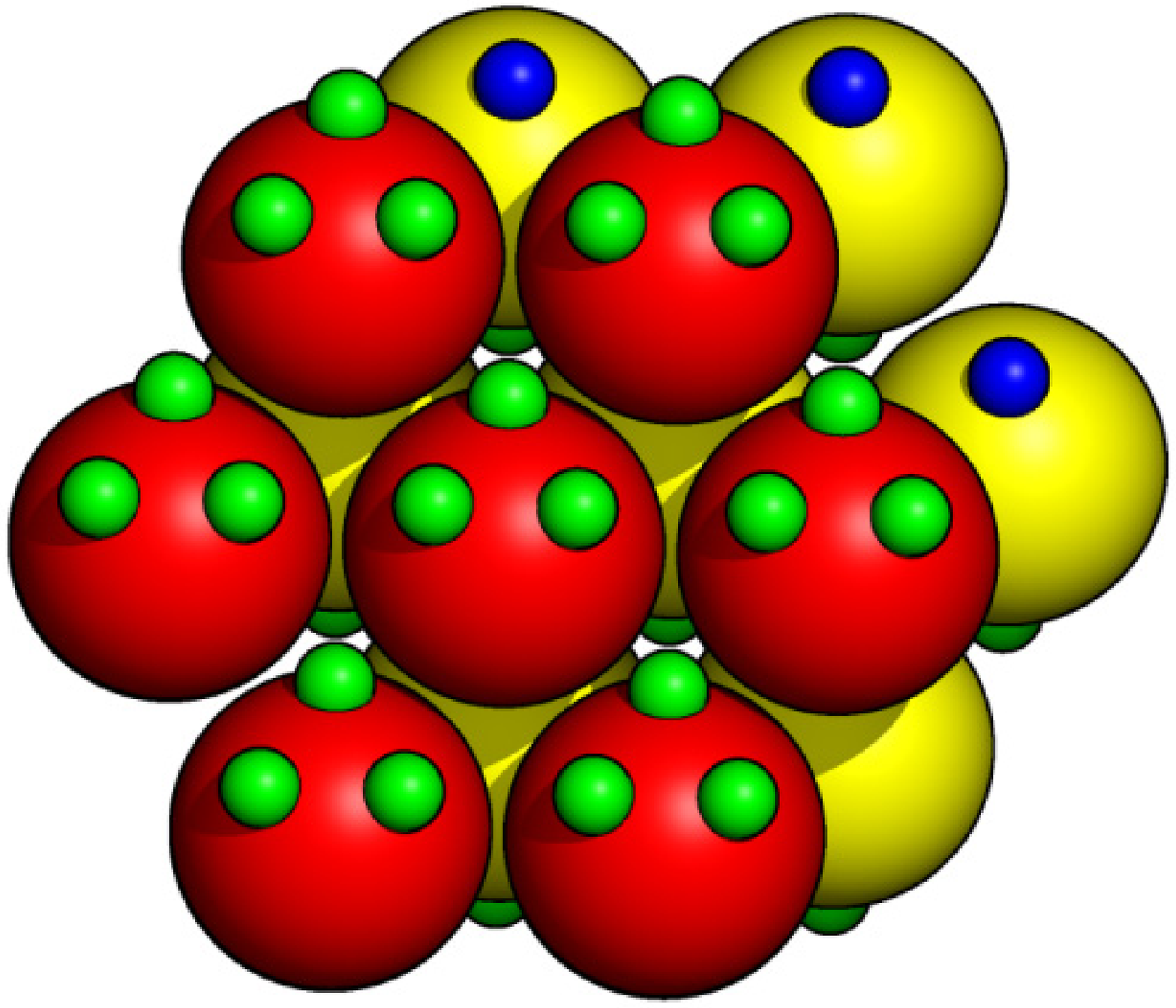}} 
\parbox[c]{6.1cm}{\includegraphics[width=6.0cm]{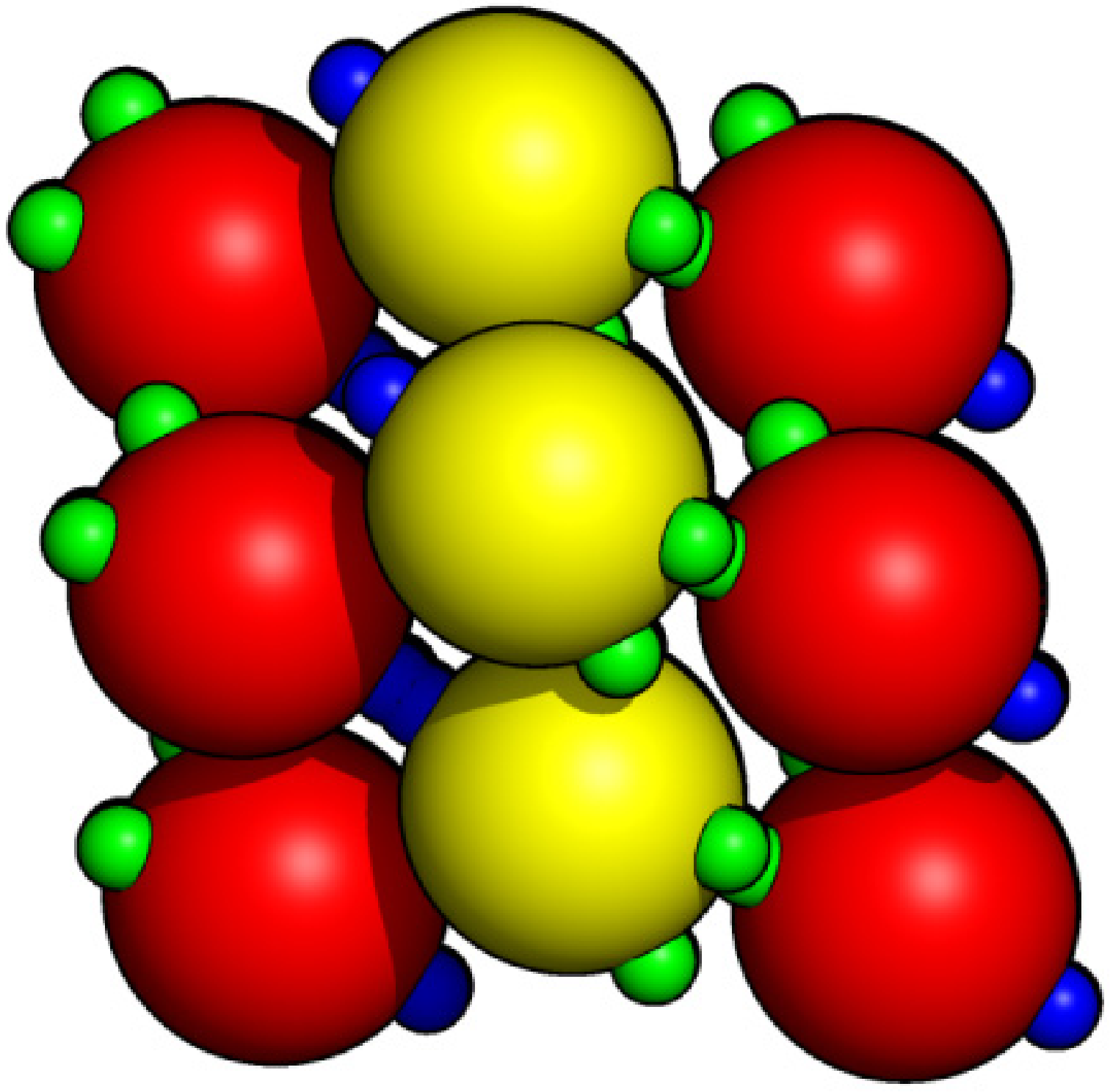}}
\end{tabular}
\end{center}
  \caption{\large Perpendicular views of the low-pressure (top) and
    high-pressure (bottom), ordered equilibrium structure obtained for
    $g = 150.00$ (labels ``g - hexagonal layers III'' and ``h - hexagonal packed III'' in Figure
    2 of the main article). The colour code for the blue and
    green patches has been specified in Figure 1 of the main article, the
    colours red and yellow for the patchy particles have been
    introduced for convenience$^\ast$.}
  \label{fig:g_150}
\end{figure}

\end{document}